\documentclass[12pt]{article}

\usepackage{booktabs}
\usepackage[english]{babel}
\usepackage{amsmath,amssymb,amsbsy,amstext, amsthm, simplewick}
\usepackage{graphicx}
\usepackage{amsfonts}
\usepackage{amssymb}
\usepackage{upgreek}
\usepackage{simplewick}
\usepackage{exscale,relsize}
\usepackage{bbding}
\usepackage{cite}
\usepackage{tabu}
\usepackage{multirow}

\usepackage{sectsty,color}
\usepackage{graphicx}  
\usepackage{dcolumn}   
\usepackage{bm}        
\usepackage{amssymb}   
\usepackage{amsmath} 
\usepackage{enumerate}
\usepackage{graphicx}  
\usepackage{dcolumn}   
\usepackage{bm}        
\usepackage{amssymb}   
\usepackage{amsmath} 
\usepackage{enumerate}

\newcommand{\Expect}[1]{\left\langle #1 \right\rangle}

\newcommand{\bea}{\begin{eqnarray}}
\newcommand{\eea}{\end{eqnarray}}

\newcommand{\beq}{\begin{equation}}
\newcommand{\eeq}{\end{equation}}
\newcommand{\be}{\begin{equation}}
\newcommand{\ee}{\end{equation}}

\def\gev{\, {\rm GeV}}

\def\cm{\, {\rm cm}}

\newcommand{\gsim}{\lower.7ex\hbox{$\;\stackrel{\textstyle>}{\sim}\;$}}
\newcommand{\lsim}{\lower.7ex\hbox{$\;\stackrel{\textstyle<}{\sim}\;$}}

\begin{document}

\thispagestyle{empty}

\vspace*{1.31cm}

\begin{center}
{\LARGE \bf {Bounds on self-interacting fermion dark matter from observations of old neutron stars}}\\

\vspace*{1.61cm} {\large
Joseph Bramante$^{\bigstar,\lozenge}$\footnote{\tt
josephbramante@gmail.com},
Keita Fukushima$^{\bigstar}$\footnote{\tt keitaf@hawaii.edu},
Jason Kumar$^{\bigstar}$\footnote{\tt jkumar@hawaii.edu}, and Elan Stopnitzky$^{\bigstar}$\footnote{\tt stopnitzky@hawaii.edu}}\\
\vspace{.5cm}
{$^{\bigstar}$ Department of Physics and Astronomy,
University of Hawaii}\\
{ $^{\lozenge}$ Department of Physics,
University of Notre Dame}\\

\vspace{1.5cm}

{\Large ABSTRACT}
\end{center}

{The existence of old neutron stars deeply constrains self-interacting fermion dark matter, which can form star-killing black holes. We quantify this constraint on dark matter-nucleon scattering, considering collapse scenarios that broaden bounds over intermediate masses. We then find the self- and co-annihilation rates necessary to lift these dark matter-nucleon scattering bounds. For Yukawa-coupled dark matter that fits dwarf galaxy halo profiles with a coupling $\alpha = 10^{-1}-10^{-4}$, a scalar mediator mass $m_\phi = 1-500$ MeV, and DM mass $m_X = 0.1-10^7$ GeV, we show that fermion dark matter is unconstrained if it self-annihilates at a rate greater than $10^{-40} ~ \rm{cm^3/s}$ or co-annihilates with baryons at a rate greater than $10^{-50} ~ \rm{cm^3/s}$.}
\vspace{2cm}


\setcounter{page}{0} \setcounter{footnote}{0}

\section{Introduction}

In this work, we study constraints on self-interacting fermionic dark matter (DM) which arise from observations of
neutron stars.  There is a lively debate within the astrophysics community regarding whether or not data
on matter distributions favor collisionless dark matter or self-interacting dark matter (SIDM) \cite{Spergel:1999mh,Vogelsberger:2012ku,Rocha:2012jg,Peter:2012jh,Klypin:1999uc,Koposov:2009ru,Oh:2010mc,Sawala:2010zw,BoylanKolchin:2011de,BoylanKolchin:2011dk,Pontzen:2011ty,Governato:2012fa,Walker:2011zu,Zavala:2012us,Kahlhoefer:2013dca}.
Although there has been no definitive resolution to this question, there has been heightened interest in
constraints on self-interacting dark matter.
In this work, we will consider astrophysical constraints on a particular type of self-interacting dark matter: fermionic DM with an attractive self-interaction potential and a small annihilation cross
section.  This class of dark matter can be constrained by the observation of old neutron stars.

The principle behind this bound is
simple~\cite{Kouvaris:2010vv,Kouvaris:2010jy,deLavallaz:2010wp,McDermott:2011jp,Kouvaris:2011fi,Guver:2012ba,Kouvaris:2012dz,Bramante:2013hn,Bell:2013xk,Jamison:2013yya,Bertoni:2013bsa,Kouvaris:2013awa,Kouvaris:2011gb}:
when a dark matter particle scatters with nucleons in a neutron star it will lose kinetic energy to elastic scattering, and if the dark matter
falls below the escape velocity then it will be gravitationally captured and will  collect at the core of the neutron star.
Assuming the resulting agglomeration of dark matter is not depleted by annihilation, the dark matter will simply keep collecting
until it collapses into a black hole.  If the black hole can efficiently consume the neutron star within its lifetime, then we should not see
older neutron stars in dense pockets of dark matter.  The observation of such neutron stars then effectively constrains the dark matter capture rate, and
in turn the dark matter-neutron scattering cross section.

This type of analysis has been used largely for asymmetric dark matter \cite{Nussinov:1985xr,Hooper:2004dc,Gudnason:2006ug,Khlopov:2007ic,Foadi:2008qv,Kaplan:2009ag,Cohen:2010kn,Shelton:2010ta,Blennow:2010qp,Davoudiasl:2010am,Buckley:2010ui,Falkowski:2011xh,Hall:2010jx,Haba:2010bm,Frandsen:2011kt,Graesser:2011wi,Cheung:2011if,MarchRussell:2011fi,Dutta:2010va,Gu:2010ft,Allahverdi:2010rh,DelNobile:2011je,Tulin:2012re,Iminniyaz:2011yp,Buckley:2011ye,Buckley:2011kk,Cui:2011qe,Davoudiasl:2011fj,Lin:2011gj,Kumar:2011np,Kaplan:2011yj,Cirelli:2011ac,MarchRussell:2012hi,Unwin:2012rp,Choi:2012ba,Casanellas:2012jp,Perez:2013nra,Pearce:2013ola,Gelmini:2013awa,Okada:2013cba,Petraki:2013wwa,Cheung:2013dca,Bhattacherjee:2013jca,Zhang:2013ama,Casanellas:2013nra,Zurek:2013wia,Iminniyaz:2013cla,Kumar:2013vba}, considered the prototypical
example of non-annihilating dark matter.
But in fact, it is not necessarily true that asymmetric dark matter cannot self-annihilate.  The usual assumption is that the dark particle can
only annihilate against the dark anti-particle, which is not abundant in nature.  But particle/particle self-annihilation is only forbidden
if the dark matter is the lightest particle charged under a phase rotation symmetry~\cite{Bramante:2013hn,Kumar:2013vba}.  If the dark matter is stabilized by a parity
symmetry, then even asymmetric dark matter can self-annihilate.  More generally, then, we see that observations of old neutron stars
provide bounds on any dark matter model with very weak self-annihilation, and the bounds are a function of the self-annihilation
cross section.
Indeed, it has been shown that even a self-annihilation cross section as small as $\sim 10^{-41}~\rm{cm^3/s}$ is sufficient to eliminate any constraints
on bosonic dark matter~\cite{Bramante:2013hn} (assuming an ambient dark matter density of $\sim 0.3~\gev / \cm^3$).

Neutron star constraints have mostly been applied to bosonic dark matter, which has no Fermi
degeneracy pressure to obstruct gravitational collapse.  In the absence of self-interactions, the bosonic Chandrasekhar
bound scales as $N_{Chand.} \propto (m_{pl} / m_X)^2$, where $m_X$ is the dark matter mass.  For non-interacting
fermions, the Chandrasekhar bound instead scales as $N_{Chand.} \propto (m_{pl} / m_X)^3$ due to the degeneracy
pressure.  As a result, old neutron stars could not have accumulated enough non-interacting fermionic dark matter to
initiate gravitational collapse.  If bosonic dark matter has even a very small repulsive self-interaction,
then the bosonic Chandrasekhar bound could also scale as $(m_{pl} / m_X)^3$, leading to elimination of any
bound on dark matter.

However, if fermion dark matter has an attractive self-interaction, then this force can compensate
for the Fermi degeneracy pressure, allowing for dark matter collapse to black holes.  For bosonic dark matter,
a purely attractive interaction may be expected to lead to a vacuum instability.  But for fermionic
dark matter, a Yukawa interaction could lead to a consistent attractive self-interaction.  It is
thus of interest to determine the extent to which fermionic dark matter with Yukawa self-interactions
can be constrained by observations of old neutron stars, and in particular, if the constraints have
implications for the parameter space relevant for astronomical evidence for SIDM. Here we will demonstrate
that an attractive self-interaction $\sigma / m_X \sim 10^{-24} ~\rm{cm^2/GeV}$ sufficient to fit halo profiles
of dwarf galaxies usually implies deep constraints on the dark matter-neutron scattering cross section ($\sigma_{n X} < 10^{-45}~\rm{cm^2}$).
In order to lift this bound, dark matter in this phase space must have a self- or co-annihilation interaction.

Neutron star constraints
on non-annihilating fermion dark matter with Yukawa interactions have been previously considered in \cite{Kouvaris:2011gb},
and it was found that, for the range of dark matter self-interaction parameters relevant for cosmology,
neutron star observations could be highly constraining.  In this work, we find large regions of $(m_X, \sigma_{nX})$
parameter space which can be probed by direct detection experiments, but would be inconsistent with neutron
star observations if dark matter were non-annihilating.
We then address a complementary question: what are the implications of a small self-annihilation or
co-annihilation cross section on these constraints from neutron star observations, and in particular, how large
would these annihilation cross sections have to be in order to allow consistency between neutron star observations and potential
direct detection signals at upcoming experiments?  A scenario relevant to this analysis would be the case of asymmetric
dark matter which is stabilized by a continuous phase rotation symmetry which is very weakly broken.  If the
stabilizing symmetry is weakly broken to a parity, then a very small self-annihilation cross section would
be allowed, and if it is weakly broken altogether then a very small dark matter-neutron
co-annihilation cross section would also be allowed.  Another example would be oscillating asymmetric dark matter \cite{Buckley:2011ye,Tulin:2012re}.

Here we summarize the analysis of this paper:
\begin{itemize}
\item[I.] {\it Accumulated DM fermions.}  The number of dark matter fermions collected in the neutron star depends on the rate of capture ($C_X$), local dark matter density ($\rho_X$),
the fermion-neutron scattering cross section ($\sigma_{nX}$), and the dark matter mass ($m_X$).
Dark matter in the neutron star can be depleted by self-annihilation or co-annihilation.
\item[II.] {\it Collapse via Yukawa coupling.}  Fermionic dark matter particles will form a black hole if the energy of collected fermions is minimized
for a radius smaller than the Schwarzschild radius. For the density and number of DM fermions achievable in neutron stars, this requires enough attractive
coupling between fermions to overcome Fermi degeneracy pressure.
\begin{itemize}
\item[a.] The strength ($\alpha$) and mediator mass ($m_\phi$) of an attractive Yukawa coupling ($V_{\rm Yuk.} =\sum_{j} \alpha e^{-m_\phi r_j}$) will
determine whether the fermions are degenerate or non-degenerate, and relativistic or non-relativistic, when collapse occurs in a neutron star.
\item[b.] Collapse from a non-degenerate state generally requires heavier
(and more to the point fewer) dark matter particles. If the DM collapses while non-degenerate, it will become degenerate during collapse and must have
a strong enough self-attraction to continue collapsing in spite of degeneracy pressure.
\end{itemize}
\item[III.] {\it Destruction of the neutron star.} If the rate at which the black hole accretes nearby baryons outstrips Hawking radiation, the black
hole will destroy the neutron star.
\item[IV.] {\it Constraints on self-interacting dark matter.} Some models of asymmetric, fermionic dark matter with an attractive self-interaction tuned to galactic core
profiles \cite{Tulin:2012wi,Tulin:2013teo} are prohibited by old neutron stars. However, we show more generally that any fermionic dark matter with a Yukawa-coupling
in the range $10^{-4}<\alpha<10^{-1}$ will evade these neutron star bounds with a self-annihilation cross section
greater than $10^{-40} \rm ~ cm^3/s$ or a cross section for co-annihilation with neutrons greater than $10^{-50} \rm ~ cm^3/s$.
\end{itemize}

In Section \ref{secacc} we detail dark matter fermion accumulation in neutron stars for both self-annihilating and co-annihilating DM fermions. We find
the number of dark matter particles which must accumulate in non-degenerate and degenerate states in order to form a black hole and destroy the neutron star in
Section \ref{sectioncollapse}. In Section \ref{sectionconstraints}, we determine the bounds old neutron
stars place on self-annihilating or co-annihilating dark
matter fermions with an attractive Yukawa self-interaction. In Section \ref{secconclusion}, we conclude that very small self-annihilation and
co-annihilation terms are sufficient to lift neutron star constraints on Yukawa-coupled dark matter fermions.

\section{Accumulation of dark matter in a neutron star} \label{secacc}
If fermionic dark matter scatters with nucleons, dark matter in the vicinity of neutron stars will be deflected into the gravitational potential
of the neutron star and repeatedly scatter into a thermalized core at the center of the neutron star. For this study we consider fermionic
dark matter with a Yukawa coupling which either self-annihilate or co-annihilate with neutrons.

In order for a scattering interaction to occur, a dark matter particle must impart enough momentum to a degenerate neutron for it to exceed the
Fermi-momentum $p_{F}$. Thus, the fraction of neutrons that are able to participate in capture will depend on the dark matter momentum.
If the average dark matter velocity, escape velocity, and baryon density are assumed to be uniform within the neutron star, then the capture
rate $C_X$ is given by \cite{Kouvaris:2010jy,McDermott:2011jp,Kouvaris:2007ay}
\begin{align}
C_X \simeq \sqrt{\frac{6}{\pi}}\frac{\rho_{X}}{m_{X}}\frac{v(r)^{2}}{\bar{v}}f(\sigma_{nX})\xi N_{B}\left[1-\frac{1-\exp(-B^{2})}{B^{2}}\right]
\end{align}
where $N_{B}$ is
the total number of neutrons within the star, $\rho_X$
is the ambient density of dark matter, $m_X$ is the mass of the dark matter, $v(r)$ is the infalling dark matter speed which is around the escape velocity at the star's surface $v(r) \sim v_{esc} \simeq 1.8
\times 10^5$, $\bar{v}$ is the average
dark matter speed in the rest frame of the neutron star, $f(\sigma_{nX})$ is a function of the DM-neutron scattering cross section which
gives the probability that an incident dark matter particle will scatter, and
\begin{align}
B^{2} = \frac{6 \, v_{esc}^{2}}{\bar{v}^2}\frac{m_{X} m_{B}}{(m_{X}-m_{B})^{2}},
\end{align}
accounts for the minimum energy loss necessary to capture the dark matter, where $m_B = m_n$ is the mass of the scattering baryon.
Here, we assume that $\sigma_{nX}$ is velocity-independent.

$\xi$ is the fraction of neutrons which contribute to scattering after accounting for Pauli blocking, that is, after accounting for the
fact that dark matter cannot scatter off neutrons which are so far below the Fermi surface that it is kinematically
impossible for them to be excited above the Fermi surface.
Dark matter particles entering the star will attain speed $v(r)$ which is greater than the escape velocity $v_{esc}$.
For typical neutron star parameters $\rho_{B}=7.8\times 10^{38} \, $GeV$/ $cm$^{3}$ and $v_{esc} \simeq 1.8
\times 10^5$ km/s at the surface, we then find $\xi = 1$ for all $m_{X} \gtrsim 1\, $GeV. For smaller
values of $m_{X}$, $\xi \sim (\sqrt{2} m_r \, v_{esc}/p_{F})$, where $p_{F} \simeq \left(3
\pi^{2}\rho_{B}/m_{B}\right)^{\frac{1}{3}} $ is the Fermi level and $m_r = m_X m_B/(m_X + m_B)$ is
the dark matter-neutron reduced mass.

As expected, the capture rate increases with the cross section $\sigma_{nX}$, since
$f (\sigma_{nX}) = \sigma_{sat.} \times \\ \left(1 - \exp (-\sigma_{nX} / \sigma_{sat.})\right)$. But the likelihood of an incident dark matter particle
scattering saturates to 1 for $\sigma_{nX} > \sigma_{sat.} \sim R_{n}^{2}/N_B$.
Note that for low mass
dark matter, Pauli-blocking decreases the effective number of neutrons available for scatter, increasing $\sigma_{sat}$.
In full detail \cite{Petraki:2013wwa,Kouvaris:2007ay} it has been shown that
the scattering cross section saturates when $\sigma_{sat} \simeq R_{n}^{2}/(0.45N_{B} \xi)$.
For higher mass dark matter and fiducial values $M_{ns} = 1.44 \, M_{\bigodot}$ and $R_{n}=10 ~ $km, the maximum capture
cross section is $\sigma_{sat} \simeq 10^{-45} ~{\rm cm^{2}}$.

The rate at which dark matter accumulates in the star will increase with $C_{X}$ and be suppressed by evaporation, decay, and
self-annihilation. Halo formation and CMB observations constrain the characteristic decay lifetime to the range $\tau_{X} > 10 \, $Gyr$ $,
making this channel an insignificant source of dark matter depletion
over the neutron star lifetime $t_{ns}$ \cite{Bramante:2013hn,Peter:2010jy,Aoyama:2011ba}.
Evaporation has a negligible effect for a typical neutron star if $m_{X} \gsim 48~{\rm eV}$~\cite{McDermott:2011jp}.
Thus observationally permitted models of decaying, self-annihilating, and self-interacting dark matter
will collect in a neutron star at a simple rate \cite{Bramante:2013hn,Petraki:2013wwa,Kouvaris:2007ay,Bell:2013xk}:
\begin{align}
\frac{dN_{\rm acc}}{dt} \approx C_{X} - \frac{\Expect{\sigma_a v} N_{\rm acc}^{2}}{V_{th}}
\end{align}
where the second term gives the rate of self-annihilation, $\Expect{\sigma_a v}$ is the annihilation cross section,
and $V_{th} = \frac{4}{3} \pi r_{th}^3$ is the volume within which the dark matter collects. The thermalization radius $r_{th}$ is
determined in
Section \ref{sectioncollapse} through application of the virial theorem. Accordingly, its value will depend on the kinetic
energy of the dark matter particles, which in turn depends on whether they are non-degenerate or degenerate. This leads to two limits
for the self-annihilation rate:

\begin{itemize}
\item[I.] For non-degenerate dark matter particles, (see Eq. \eqref{rthermclass}) $r_{th} = \left( \frac{9 k_B T}{4 \pi G \rho_b
m_X}\right)^{1/2} $ and is independent of $N_{acc \rm}$. Eq. (3) can then be solved straightforwardly to give

\begin{align}
N_{\rm acc} (t) \approx \sqrt{\frac{C_X V_{th}}{\langle \sigma_a v \rangle} }
{\rm Tanh} \left[ \sqrt{\frac{C_X \langle \sigma_a v \rangle}{V_{th}}} t \right],
\qquad (t \leq t_{non-deg.}) \label{naccfree}
\end{align}
where $t_{non-deg.}$ is the amount of time that elapses before the fermions become degenerate.
To find $t_{non-deg.}$, we substitute
the number of fermions required for degeneracy (Eq. \eqref{ndeg}) into \eqref{naccfree} and solve numerically for $t_{non-deg.}$.
Of course if the self-annihilation rate is large enough, the fermions will never be in a degenerate state.

\item[II.]
For some regions of fermion
dark matter parameter space, the dark matter will begin collapsing before becoming degenerate.
But if the dark matter becomes degenerate, the volume it occupies will depend on $N_{\rm acc}$.
For degenerate dark matter particles, (see Eq. \eqref{rthermdeg})
$r_{th} \simeq \frac{(9 \pi N_{acc}/4)^{1/6}}{(4 \pi G \rho_b m_X^2/3)^{1/4}}$ and we see that
a non-linear differential equation governs the accumulation of annihilating dark matter fermions
into a degenerate state,
\begin{align}
\frac{dN_{ \rm acc}}{dt}
\approx C_{X} - \frac{\sqrt{2}\Expect{\sigma_a v} \left(G \rho_b m_X^2\right)^{3/4} N_{acc}^{3/2}}{(3 \pi)^{3/4}}.
\qquad ({t \geq t_{non.deg}})
\label{naccdeg}
\end{align}
\end{itemize}
We solve \eqref{naccdeg} numerically, subject to the boundary condition on $N_{\rm acc}(t_{non-deg})$ obtained by
evaluating eq.~(\ref{naccfree}) at $t_{non-deg.}$,
to determine the the total number of accumulated dark matter particles at the current
time, $N_{\rm acc}(t_{ns})$, where $t_{ns}$ is the age of the neutron star.

Of course, if the self-annihilation cross section is too large, any dark matter asymmetry will be washed out in
the early universe.  If we assume the presence of enough asymmetric dark matter to account for the cosmological
dark matter abundance with no symmetric component, then the self-annihilation process will be frozen out below
temperature $T$ provided
\bea
\langle \sigma_A v \rangle \lsim \left[ (0.2) \sqrt{g_*(T) \over 10} {m_X \over T} \right] {\rm pb}
\eea
where $g_*$ is the number of relativistic degrees of freedom at temperature $T$.  For the range of
$m_X$ and $T$ which we will consider, self-annihilation will not wash out the asymmetry provided
$\langle \sigma_A \rangle \ll {\rm pb}$.

Besides self-annihilation, co-annihilation of dark matter with baryons can deplete dark matter in the neutron star.
For example, if the symmetry which stabilizes the dark matter is weakly broken, then the process $nX \rightarrow n+{anything}$
is allowed (if baryon number can be violated, then the final state need not have a neutron).  The total rate
of co-annihilations, like a total decay rate, is proportional to $N_{acc}$.  Assuming no self-annihilations, the number of
accumulated dark matter particles is then a solution to the following differential equation: $\frac{d N_{acc} }{dt} = C_X - N_{acc}/\tau$,
where $1/\tau$ is some decay or co-annihilation rate.

Unlike a co-annihilation rate, a decay cannot significantly suppress the rate of dark matter accumulation in neutron stars \cite{Bramante:2013hn}, because of stringent astrophysical and cosmological bounds on dark matter decay. A decay rate large enough to significantly impact the amount of accumulated dark matter over the lifetime of a neutron star would also deplete the cosmological abundance of dark matter by an unacceptable amount and disrupt observed halo profiles \cite{Peter:2010jy,Aoyama:2011ba}. On the other hand, a co-annihilation rate can depend on the baryon density; it is thus possible for there to be a large co-annihilation rate within a neutron star, but an insignificant rate anywhere else.

The accumulation rate for dark matter co-annihilating with baryons is
\begin{align}
N_{\rm acc, coann.} = \frac{C_X}{  n_{B} \Expect{\sigma_a v}_{co} } \left(1-e^{-t_{ns} n_{B} \Expect{\sigma_a v}_{co} }\right), \label{coan}
\end{align}
where $\Expect{\sigma_a v}_{co}$ is the co-annihilation cross section and $n_{B}$ is the number density of baryons in the neutron star.
Co-annihilation with baryons is a well-motivated feature of asymmetric dark matter models, which relate the baryon asymmetry to a
dark sector asymmetry. However the required level of co-annihilation to lift neutron star bounds is typically smaller than the
required self-annihilation \cite{Bramante:2013hn,Bell:2013xk}, because a neutron star provides a constant, extremely dense background
of baryons, while in the case of dark matter self-annihilation, a substantial quantity of dark matter must collect for efficient annihilation.

\section{Collapse of Yukawa-coupled fermions} \label{sectioncollapse}

Here we consider the collapse of a collection of dark matter fermions thermalized within the core of a neutron star.
This analysis largely follows that of~\cite{Kouvaris:2011gb}, but with more detail, and with some new and more precise
statements.  We also follow the notation of~\cite{Kouvaris:2011gb}.
To find the number of dark fermions necessary to initiate collapse, we use the virial theorem
and determine the number of dark fermions ($N_{\rm coll.}$) necessary to minimize the energy at an arbitrarily small radius.
The fullest statement of the virial theorem
relates the time-averaged kinetic energy of a system of particles to the time-averaged sum of forces acting on each particle,
$
\Expect{\sum_i \vec{p_i} \cdot \vec{v_i}} = - \Expect{ \sum_i \vec{F_i} \cdot \vec{r_i} }. \label{virialprecise}
$
Here the summation over $i$ adds the forces on each particle while $\vec{r_i}$ is each particle's position. In an isotropic potential, the
equation simplifies considerably,
\begin{align}
\sum_i \Expect{{p_i^2 \over \sqrt{p_i^2+ m^2}} } =  \Expect{ \sum_i r_i \frac{\partial V_i(r_i)}{\partial r_i} }.  \label{virialsimpler}
\end{align}
Suppose this collection of dark matter fermions has a Yukawa interaction $\mathcal{L}_X \supset  \alpha \phi \bar{X}X$ where $X$ is a dark fermion that scatters with nucleons, $\phi$ is a scalar boson mediating the Yukawa force and $\alpha$ is the Yukawa coupling. The  potential energy of a single dark matter particle $X$ at a radius $r$ from the center of a neutron star is given by
\begin{align}
V_X(r) = \frac{2 \pi G \rho_b m_X r^2 }{3} - \frac{G N_X m_X^2}{r} - \sum_j^{N_X-1} \frac{\alpha e^{-m_\phi r_j}}{r_j},
\end{align}
where the first term is the gravitational potential sourced by baryons in the neutron star, the second term is the gravitational potential sourced by dark matter fermions, and the last term is the Yukawa potential between two fermions summed over dark matter fermions in the neutron star. $N_X$ is the number of dark matter particles in the neutron star, $\rho_b$ the mass density of the neutron star at its core, $m_\phi$ the mass of the boson $\phi$, and $r_j$ the inter-particle distance of a test dark matter particle to another particle $j$. Applying the virial equation to a typical Yukawa-coupled fermion in a sphere of dark matter fermions at the core of a neutron star, we can relate the energy of the captured fermion to its potential,
\begin{align}
-{p^2 \over \sqrt{p^2 +m^2}}  + \frac{4 \pi G \rho_b m_X r^2}{3} + \frac{G N_X m_X^2}{r}
+ \sum_j^{N_X-1} \left( \frac{\alpha e^{-m_\phi r_j}}{r_j} + \alpha m_\phi e^{-m_\phi r_j} \right)=0 \label{virialfull}
\end{align}
In \eqref{virialfull} we leave the time-average implicit and identify $r$ as the radius of the fermions from the center of the neutron star.

\subsection{State before collapse}

In order to analyze the collapse of dark matter fermions, we must first determine the number of dark matter fermions and their radius
before the onset of collapse. If the number of collected dark matter fermions is small, then it is easy to estimate the dark matter
thermal radius because the Yukawa coupling and self-gravity contributions are negligible.
In the limit where the first two terms of \eqref{virialfull} dominate, we are essentially considering a three-dimensional
harmonic oscillator with frequency $\omega = [(4\pi /3) G \rho_b]^{1/2} \sim 10^{-12} (k_B T)$.  The total energy of a
single dark matter fermion (in the non-degenerate limit) would be $\sim 3 k_B T$, and the kinetic energy (from the virial theorem)
would be $\sim (3/2) k_B T$.

When enough dark matter particles accumulate ($N_{acc} \sim (k_B T / \omega)^3 \sim 10^{36}$), effects of Pauli
exclusion cannot be ignored and one would have a degenerate harmonic oscillator, with all of the lowest-lying states filled.
Note that, for all values of $m_X$ which we will consider, dark matter will still be non-relativistic at the time when it
first becomes degenerate.
We can then approximate the momentum as $p \sim N_X^{1/3}/r$ and we find that the kinetic energy can be
written as
\begin{align}
E_{k,{\rm deg.}} = \frac{(9 \pi N_X/4)^{2/3}}{2 m_X r^2},~~~~~ ~~ E_{k,{\rm non-deg.}} = \frac{3}{2} k_B T . \label{ek}
\end{align}
The pre-collapse thermal radius for dark matter fermions (degenerate or non-degenerate) is,
\begin{align}
r_{th,{\rm deg.}} &= \frac{(9 \pi N_X/4)^{1/6}}{(4 \pi G \rho_b m_X^2/3)^{1/4}} \simeq 2.4 \times 10^{-4} ~{\rm cm} \times N_X^{1/6}\left(\frac{\rm GeV}{m_X}\right)^{1/2} \label{rthermdeg}, \\
r_{th,{\rm non-deg.}} &= \left( \frac{9 k_B T}{4 \pi G \rho_b m_X}\right)^{1/2} \simeq 250 ~ {\rm cm}\times \left(\frac{\rm GeV}{m_X}\right)^{1/2} \label{rthermclass}.
\end{align}
Note that in \eqref{ek} and hereafter we assume that the dark matter fermions are two-fold degenerate.
For an $N_q$-fold degeneracy, fermions in a sphere of radius $r$ will be fully degenerate when enough particles have collected:
\begin{align}
N_{\rm deg.} = \frac{N_q  \frac{4}{3} \pi r^3 (2 \pi m_X k_B T)^{3/2}}{({2 \pi})^3}
= 2.7 \times 10^{28}N_q  \left( {r \over {\rm cm}}\right)^3 \left( {m_X \over {\rm GeV} }\right)^{3/2}  , \label{ndeg}
\end{align}
where the final equality assumes a typical neutron star temperature of $T=10^5 ~{\rm K}$.

We have thus far assumed that the self-gravity and Yukawa potential terms are not significant.
The self-gravity term is important if
\bea
N_{acc} &\geq& {4 \pi \over 3} {\rho_b r^3 \over m_X}.
\eea
For the models we are interested in, collapse will not occur until the Yukawa potential terms are large enough to overcome the Fermi degeneracy pressure, at which point the approximation of the potential as that of a harmonic oscillator is no longer valid.  If dark matter becomes degenerate only after the Yukawa terms become important, then the above estimates of the radius of the dark matter will be altered.

\subsection{Critical number for collapse}

Gravitational collapse will begin when $N$ is so large that the force due to the total potential cannot be resisted by the pressure due to
the kinetic energy.  At this point there will no longer be an equilibrium solution to the virial equation (Eq.~\eqref{virialfull}).
The kinetic energy term of Eq.~\eqref{virialfull} becomes increasingly negative as $r$ decreases.  The baryonic
gravity term changes in the same direction, becoming less positive as $r$ decreases.  We thus see that
collapse will occur when the self-gravity and Yukawa terms exceeds the kinetic energy term.
If collapse begins from a non-degenerate state, one must ensure that these conditions are also satisfied once the
fermions become degenerate. Likewise collapse beginning from a non-relativistic state must continue
when the particles become relativistic. While the intrepid reader is encouraged to proceed through the details, a tabulation of
collapse channels and corresponding phase space inequalities is included at the beginning of Section \ref{sectionconstraints}.

In order to determine the values of $r$ and $N_X$ for which collapse will occur, it is useful to simplify the Yukawa terms of Eq.~\eqref{virialfull} by re-expressing the inter-particle
distance $r_j$ as a function of $r$ and $N_X$. To that end, we find that $x \equiv (4\pi /3)^{1/3}r / N_X^{1/3} \simeq 1.6 r/N_X^{1/3}$ is the shortest inter-particle
distance in a sphere of evenly distributed dark matter fermions. This facilitates reframing Eq.~\eqref{virialfull} in terms of only $r$ and $N_X$
for two limits: the ``strongly-screened" limit, for which the Yukawa potential is sourced only
by nearest neighbors of the dark matter particle, and
the ``partly-screened" limit, for which the Yukawa potential can be approximated as a Coulomb potential sourced by all unscreened neighbor particles. For a thorough derivation
of the virial equation in the strongly-screened limit ($m_\phi x \gtrsim 1$) and partly-screened limit ($m_\phi x \lesssim 1$), see Appendix \ref{appyukawa}. Here we quote the
results. In the strongly-screened non-relativistic limit ($m_\phi x \gtrsim 1$), Eq.~\eqref{virialfull} becomes
\begin{align}
-2 E_k + \frac{(\frac{4}{3} \pi)^{1/3} G N_X^{2/3} \rho_b  m_X y^2}{m_\phi^2} + \frac{(\frac{4}{3} \pi)^{1/3} G N_X^{2/3} m_X^2 m_\phi}{y}  + 8\alpha \left( \frac{m_\phi e^{-y}}{y} + m_\phi e^{-y} \right)=0 \label{virialscreened}.
\end{align}
and in the partly-screened non-relativistic limit ($m_\phi x < 1$),
\begin{align}
-2 E_k + \frac{(\frac{4}{3} \pi)^{1/3} G N_X^{2/3} \rho_b  m_X y^2}{m_\phi^2} + \frac{(\frac{4}{3} \pi)^{1/3} G N_X^{2/3} m_X^2 m_\phi}{y}  + \frac{4 \pi \alpha m_\phi}{y^3}=0 \label{virialps},
\end{align}
where $y$ is a convenient dimensionless variable, the argument of the Yukawa exponential for nearest neighbor fermions, $y \equiv x m_\phi \simeq 1.6 m_\phi r / N_X^{1/3}$.

We can now determine the critical number of dark matter fermions required for collapse in four relevant regions of phase space -- degenerate or non-degenerate and
either strongly-screened or partly-screened (for now we confine our attention to non-relativistic collapse).
\begin{itemize}
\item[1.] \textit{Degenerate and partly-screened.} The dark matter fermions may begin to collapse when they are partly-screened and degenerate. To ensure this analysis remains in phase space where degenerate collapse \textit{begins} only after the dark matter is partly-screened (i.e.~dark matter will not collapse for $y \gtrsim 1$), we compare the Yukawa terms of Eqs. \eqref{virialscreened} and \eqref{virialps} to the degenerate kinetic energy \eqref{ek}. We find that collapse begins from a degenerate, partly-screened phase only if $\alpha \lesssim m_{\phi}/m_X$. Substituting the degenerate kinetic energy \eqref{ek} into the partly-screened virial equation \eqref{virialps} and dropping the small self-gravity of the dark matter fermions, we find
\begin{align}
-\frac{(3 \pi^2)^{2/3}m_\phi^2}{m_X y^2} + \frac{(\frac{4}{3} \pi)^{1/3} G N_X^{2/3} \rho_b  m_X y^2}{m_\phi^2} + \frac{4 \pi \alpha m_\phi}{y^3}=0. \label{dpscoll}
\end{align}
Well before collapse begins in this regime, the first two terms will dominate and the dark matter fermions will occupy a sphere of
radius $r_{th,{\rm deg.}}$ \eqref{rthermdeg}.
As more dark matter collects, $y$ decreases and the attractive Yukawa potential becomes important.
Collapse begins when the Yukawa term grows larger than the degenerate kinetic term, implying that the Yukawa force is large enough to overcome the Fermi
degeneracy pressure.  This occurs for
\begin{align}
y_{\rm coll.,dps} = \frac{4\pi \alpha m_X}{(3 \pi^2)^{2/3}m_\phi}.
\end{align}
Since $y=1.6 m_\phi r / N_X^{1/3}$ and $r \leq r_{th,{\rm deg.}}$ \eqref{rthermdeg}, we find that the number of dark matter particles
needed for collapse from a degenerate partly-screened state is
\begin{align}
N_{\rm coll.,dps} = 1.1 \times 10^{61} \alpha^{-6} \left( \frac{m_\phi}{\rm GeV} \right)^{12} \left(\frac{ m_X} {\rm GeV}\right)^{-9}
\label{dpsncoll}
\end{align}
\item[2.] \textit{Non-degenerate and partly-screened.} Collapse will not begin from a non-degenerate and partly-screened state
for the parameter space of interest. For dark matter fermions to collapse from a non-degenerate and partly-screened state, $y=1.6 m_\phi r / N_X^{1/3} \lesssim 1$.
For $r=r_{th,non-deg.}$ \eqref{rthermclass}, this implies $N_X \gtrsim 10^{49} (m_\phi / {\rm GeV})^3(m_X / {\rm GeV})^{-3/2}$. Recalling the number of
dark matter fermions which
can remain non-degenerate ($\sim 10^{36}$), this region of phase space is only constrained for heavier dark matter with light mass scale mediators $\phi$.
We do not consider this region of parameter space, because for heavier dark matter and lighter mediators, the dark matter will collapse long before enough
particles have collected for the dark matter to become partly-screened in a non-degenerate regime. For non-degenerate and partly-screened fermions the virial equation is the same as for degenerate
and partly-screened fermions \eqref{dpscoll}, except the kinetic energy for non-degenerate fermions is $E_{k,{\rm non-deg.}}=(3/2) k_B T$.
For a neutron
star with temperature $T=10^5 ~K$, the number of non-degenerate and partly-screened dark matter fermions required for collapse is
\begin{align}
N_{\rm coll.,ndps} = 2 \times 10^{40} \alpha^{-1} \left( \frac{m_\phi}{\rm GeV} \right)^2
\left(\frac{ m_X} {\rm GeV}\right)^{-3/2} .
\label{fpsncoll}
\end{align}
Comparing this to the aforementioned $N_X \gtrsim 10^{49} (m_\phi / {\rm GeV})^3(m_X / {\rm GeV})^{-3/2}$, we see that for all parameter space of interest, the number required for collapse is less than the number required to attain non-degenerate, partly-screened dark matter.

\item[3.] \textit{Degenerate and strongly-screened.}
For $\alpha \gtrsim m_\phi / m_X$,
degenerate  dark matter fermions will still be strongly-screened ($y \gtrsim 1$) at the onset of collapse.
The virial equation for a strongly-screened Yukawa potential \eqref{virialscreened} and degenerate fermions is
\begin{align}
-\frac{(3 \pi^2)^{2/3}m_\phi^2}{m_X y^2} + \frac{(\frac{4}{3} \pi)^{1/3} G N_X^{2/3} \rho_b  m_X y^2}{m_\phi^2} + 8\alpha \left( \frac{m_\phi e^{-y}}{y} + m_\phi e^{-y} \right)=0, \label{dscoll}
\end{align}
where again we take the fermion self-gravity as negligible. Equating the Yukawa and kinetic energy terms of Eq. \eqref{dscoll},
one could solve numerically for $y_{\rm coll.,ds}(m_X,m_\phi,\alpha)$ at which collapse begins.
Using $r \leq r_{th, deg.}$ (eq.~\eqref{rthermdeg}), one could
then estimate the maximum number of degenerate, strongly-screened fermions necessary for collapse, $N_{\rm coll.,ds}(m_X,m_\phi,\alpha)$.

However, this procedure somewhat overestimates the collapse number, because it neglects the contribution to collapse from the baryon gravity term. A more thorough, computationally intensive procedure (which we use) is to solve \eqref{dscoll} for
$y$ for successively larger values of $N_X$, until $N_X$ is large enough
that there is no solution to \eqref{dscoll} with $y>1$. This indicates that the dark matter has collapsed to a partly-screened state, and it will continue to collapse because the partly-screened Yukawa term will be much larger than the degenerate Fermi pressure given that the strongly-screened Yukawa term at greater $y$ was larger than the Fermi pressure.

\item[4.] \textit{Non-degenerate and strongly-screened.} For non-degenerate and strongly-screened Yukawa-coupled fermions,
we again equate the kinetic and Yukawa terms in the virial equation, but this time retain the dark matter self-gravity term, which
can be substantial for the higher mass, more weakly-coupled dark matter that is constrained by collapse from a non-degenerate and strongly-screened state,
\begin{align}
-3 k_B T + \frac{(\frac{4}{3} \pi)^{1/3} G N_X^{2/3} \rho_b  m_X y^2}{m_\phi^2} + \frac{(\frac{4}{3} \pi)^{1/3} G N_X^{2/3} m_X^2 m_\phi}{y} + 8\alpha \left( \frac{m_\phi e^{-y}}{y} + m_\phi e^{-y} \right)=0, \label{fscoll}
\end{align}
After substituting the thermal radius $r=r_{th, non-deg}$ (Eq. \eqref{rthermclass}), we solve numerically for the number of non-degenerate,
strongly-screened Yukawa-coupled fermions necessary to initiate collapse, $N_{\rm coll.,nds}(m_X,m_\phi,\alpha)$.

After the dark matter fermions collapse from a non-degenerate and strongly-screened state, they will become degenerate
when they have shrunk to a small enough radius.  At the point where the dark matter becomes degenerate, it will either
still be strongly-screened, or will already have become partly screened.
We can distinguish these regimes using the quantity $y_{deg.}$,
defined by evaluating the expression for $y$ at $N_X = N_{deg}$ (eq. \eqref{ndeg}),
yielding
\begin{align}
y_{deg.} \equiv \frac{1.6 m_\phi r}{N_{deg.}^{1/3}} = 2.1 \times 10^4
\left( \frac{m_\phi}{\rm GeV} \right)
\left( \frac{m_X}{\rm GeV} \right)^{-1/2} .
\end{align}

\begin{itemize}
\item[A.] \textit{Non-degenerate and strongly-screened collapse through a partly-screened degenerate state.}
In the case that $y_{deg.}<1$, the dark matter fermions collapsing from a non-degenerate and strongly-screened state will become degenerate in a
partly-screened regime. Collapse will continue if the force from the partly-screened Yukawa potential is larger than the degenerate
Fermi pressure for $y=y_{deg}.$, which is satisfied when
\begin{align}
\alpha > 1.6 \times 10^4
\left( \frac{m_\phi}{\rm GeV} \right)^2
\left( \frac{m_X}{\rm GeV} \right)^{-3/2} .
\end{align}
\item[B.] \textit{Non-degenerate and strongly-screened collapse through a strongly-screened degenerate state.}
Less massive dark matter may satisfy $y_{deg.}>1$, in which case the dark matter remains strongly-screened even after it becomes degenerate.
In this case, collapse continues if the strongly-screened Yukawa force exceeds the degenerate Fermi pressure,\footnote{This portion of collapse-permitting parameter
space yields bounds on intermediate masses not excluded in the analysis of~\cite{Kouvaris:2011gb}.}
\begin{align}
\alpha > 2.7 \times 10^{-9} \frac{e^{2.1 \times 10^4 \left( \frac{m_\phi}{\rm GeV} \right)
 \sqrt{\frac{\rm GeV}{m_X} }}}{ \left(\frac{{m_\phi}}{ {\rm GeV} }\right)\left( 1+ 4.7 \times 10^{-5}\left( \frac{\rm GeV}{m_\phi} \right)
\sqrt{ \frac{m_X}{\rm GeV} }\right)}.
\label{eq2nduscoll}
\end{align}
\end{itemize}
\end{itemize}

\subsection{Relativistic collapse}

As the fermions continue to collapse, they may become relativistic prior to forming a black hole. For a fully degenerate Fermi gas, the momentum of two-fold degenerate fermions will be $p_{deg} = (9 \pi N/4)^{1/3}/{r}$. This momentum will become relativistic when it exceeds the rest mass of the dark matter fermions, $(9 \pi N/4)^{1/3}/{r} \simeq m_X$.
Solving for the radius at which the fermions will become relativistic $r_{\rm rel.}=(9 \pi N_{\rm coll.}/4)^{1/3}/m_X$ and comparing to the Schwarzschild radius, $r_{\rm Schwarz.} = 2 G N_{\rm coll.} m_X$, we find that the fermions will become relativistic before forming a black hole if
\begin{align}
N_{\rm coll.} < 1.7 \times 10^{57} \left( \frac{\rm GeV}{m_X}\right)^3
\end{align}
which will be true for most parameter space under consideration. For this work it will be safe to assume that the Yukawa-coupled dark matter fermions are partly-screened ($y = 1.6 r m_\phi / N^{1/3} < 1$) by the time they become relativistic.  At $r=r_{rel.}$,
we find that $y < 1$, so long as $m_\phi < 0.2 m_X$. Species of Yukawa-coupled dark matter fermions with $m_\phi > 0.2 m_X$
are excluded by astrophysical bounds from gravitational lensing studies of galactic cluster collisions \cite{Kouvaris:2011gb}.

Once collapse has begun in a degenerate state, the dark matter fermions will become partly-screened and then become relativistic. Collapse continues for relativistic dark matter
fermions if the relativistic, degenerate Fermi pressure is smaller than the relativistic, partly-screened Yukawa force
at the radius at which the sphere of dark matter fermions becomes relativistic. Substituting $r_{\rm rel.}$ into the first and third terms
of Eq. \eqref{dpscoll}, the dark matter fermions will continue to collapse once relativistic if
\begin{align}
\alpha > 4.7 \frac{ m_\phi^2}{m_X^2}.
\end{align}

\subsection{Thermalization time and neutron star collapse}
After being captured in the gravitational well of the neutron star, the dark matter will repeatedly scatter with nucleons and thermalize into a core at the center of the star. To facilitate dark matter collapse, the thermalization time must be short relative to the neutron star lifetime. This condition is given by \cite{Bramante:2013hn},
\begin{align}
\sigma_{nX} < 1.1 \times 10^{-59} ~{\rm cm^2} \left( \frac{m_X}{\rm GeV}\right)^2 \left( \frac{10^5 \rm K}{T} \right) \left( \frac{\rm Gyr}{t_{th}} \right). \label{sigmatherm}
\end{align}
where $t_{th}$ is the thermalization time. In this study we shade parameter space for $t_{th} > { \rm Gyr}$, where neutron star black hole bounds will not apply. Also, by inserting
the saturation scattering cross section for high mass dark matter ($\sigma_{sat} \sim 10^{-45} {~\rm cm^2} $) and typical neutron star
parameters ($T=10^5$ K, $t_{th} =$ Gyr), we find that dark matter heavier than $m_X \gtrsim 10^7 {~ \rm GeV}$ will not thermalize quickly
and is not bound by observations of old neutron stars.

After dark matter forms a black hole in the neutron star, the black hole will only consume the neutron star if it accumulates surrounding
baryonic matter faster than the Hawking radiation rate. The change in mass of the black hole mass is given by
\begin{align}
\frac{d M_{bh}}{d t} = \frac{4 \pi (G M_{bh})^2 \rho_b }{v_s^3}  - \frac{1}{15360 \pi (G M_{bh})^2} \label{massrate},
\end{align}
where the first term on the right side is the Bondi accretion rate for neutrons, the last term is the Hawking radiation rate, $M_{bh} = N_{\rm coll.}  m_X$ is the mass of the black hole, and $v_s/c \sim 0.1$ \cite{McDermott:2011jp} is the sound speed of the neutron star.  Note that for fermionic dark matter, the accretion
of dark matter is very inefficient and can be ignored; this is in contrast to the case of bosonic dark matter.
The condition that the Bondi rate outstrip the Hawking rate is
\begin{align}
N_{\rm coll.} > 3.4 \times 10^{36} \frac{\rm GeV}{m_X}.
\end{align}
The lightest black hole which satisfies this condition ($M_{bh} = 3.4 \times 10^{36}$ GeV) will destroy a neutron star in $2 \times 10^{-3}~ {\rm Gyr} $ \cite{Bramante:2013hn}.

\section{Neutron star bounds on self-interacting dark fermions}\label{sectionconstraints}

Self-interacting dark matter fermions are bound by the existence of old neutron stars. In this section we determine what bounds old neutron stars place on fermionic dark matter with an attractive Yukawa self-interaction, and find how this bound is lifted with the addition of dark matter self-annihilation and co-annihilation. This bound depends upon whether the dark matter begins collapse from a strongly- or partly-screened degenerate state or a strongly-screened
non-degenerate state. Additionally, dark matter must accumulate in a large enough quantity to form a large black hole which can consume
an old neutron star within its lifetime. Below we tabulate the collapse channels and applicable phase space inequalities required for neutron star bounds on dark matter fermions.

\renewcommand{\arraystretch}{1.5}
{\center
\setlength{\tabcolsep}{.15em}
\begin{tabular}
{|c|*4{c|}}
 \hline
  \multicolumn{5}{|c|}{Fermion Dark Matter Collapse Bound Channels} \\
 \hline
  State of & Degenerate, & Degenerate, & \multicolumn{2}{c|}{Non-degenerate, } \\
   initial collapse & partly-screened & strongly-screened & \multicolumn{2}{c|}{strongly-screened} \\
 \hline
  Yukawa screening & $ \alpha < m_\phi/m_X $ & $\alpha > m_\phi/m_X $ & \multicolumn{2}{c|}{-} \\
 \hline
 DM accum. \# $>$   & $N_{acc}$ (Eqs.~(\ref{naccfree},\ref{naccdeg},\ref{coan}) $>$ &
 $N_{acc}$ (Eqs.~(\ref{naccfree},\ref{naccdeg},\ref{coan})) $>$  &
 \multicolumn{2}{c|} {Eqs.(\ref{naccfree},\ref{coan}) $>$} \\
 DM collapse \#  &
 Eq. \eqref{dpsncoll} &
 Sol. Eqs. \eqref{rthermdeg},\eqref{dscoll}  & \multicolumn{2}{c|}{
 Sol. Eqs. \eqref{rthermclass},\eqref{fscoll} } \\
 \hline
  Degenerate or  &  Eq. \eqref{dpsncoll} &
 Sol. Eqs. \eqref{rthermdeg},\eqref{dscoll} & \multicolumn{2}{c|}{
 Sol. Eqs. \eqref{rthermclass},\eqref{fscoll}  } \\
 non-deg.~collapse  & $>8\times 10^{35}$ &
 $>8\times 10^{35}$ & \multicolumn{2}{c|}{$<8\times 10^{35}$ } \\
 \hline
  State of second, & - & - & Partly- & Strongly-\\
    degenerate collapse & - & - & screened & screened\\
 \hline
Yukawa screening & - & - &$ \frac{2\times 10^4 m_\phi}{(m_X {\rm GeV})^{1/2}} <1 $&$ \frac{2\times 10^4m_\phi}{(m_X {\rm GeV})^{1/2}} >1 $\\
 \hline
 Continued & - & - &$ \alpha > 1.6 \times 10^4 $&$\alpha \gtrsim 2.7 \times 10^{-9}  $\\
 collapse & - & - &$ \times \frac{ m_\phi^2}{\sqrt{m_X^{3} {\rm GeV}}} $&$\times \frac{e^{2.1\times10^4  m_\phi / \sqrt{m_X {\rm (GeV)}}}}{(m_\phi/{\rm GeV})} $\\\hline
  Relativistic collapse &  \multicolumn{4}{c|}{$\alpha > 4.7 m_\phi^2/m_X^2$} \\
 \hline
  Star consumed & \multicolumn{4}{c|}{$N_{\rm coll.} >  \frac{ 3.4 \times 10^{36} \rm GeV}{m_X}$} \\
 \hline
\end{tabular}
}
\\
\\

\renewcommand{\arraystretch}{1.0}
\begin{figure}
\caption{{\bf Self-annihilation:} Neutron star collapse bounds for self-annihilating dark matter fermions with an attractive Yukawa-coupling are shown in the $(m_X, \sigma_{nX})$ plane. A neutron star with temperature $T=10^5 ~ \rm K$, an age of $7~ \rm Gyr$, and a typical ambient dark matter density of $0.3~ \rm GeV / cm^3$ with average speed $\bar{v} = 220~ \rm km/s$ is assumed. Constraint plots for Yukawa couplings $\alpha = \{10^{-1},10^{-2} \}$ and Yukawa boson masses $m_\phi = \{1,5,10  \}~{\rm MeV}$ are displayed, and bounds on dark matter with a self-annihilation cross section of $\Expect{\sigma_a v} = \{0,10^{-50},10^{-47},10^{-45},10^{-43}\} ~{\rm cm^3/s}$ are drawn with solid, dotted, dotted-dashed, short-dashed, and long-dashed lines, respectively. Horizontal arrows indicate the range of dark matter masses fitting dwarf galaxy mass halos (figure 6 in \cite{Tulin:2013teo}), assuming an attractive Yukawa coupling and the values of $\alpha$ and $m_\phi$ indicated. The shaded grey regions at the bottom right of each plot are unbounded -- they encompass parameter space where the dark matter will not thermalize at the core of the neutron star within 1 Gyr.}
 \begin{tabular}{cccc}
 & \begin{minipage}[c]{0.4\textwidth} \center ~~$\alpha =10^{-1}$ \end{minipage} &\begin{minipage}[c]{0.4\textwidth} \center ~~ $\alpha =10^{-2}$ \end{minipage} & \\
{\multirow{4}{*}{\rotatebox[origin=c]{90}{Nucleon-DM Scattering Cross Section $\sigma_{nX}$ $\rm (cm^2)$}}}
&\begin{minipage}[c]{0.40\textwidth} \includegraphics[scale=.70]{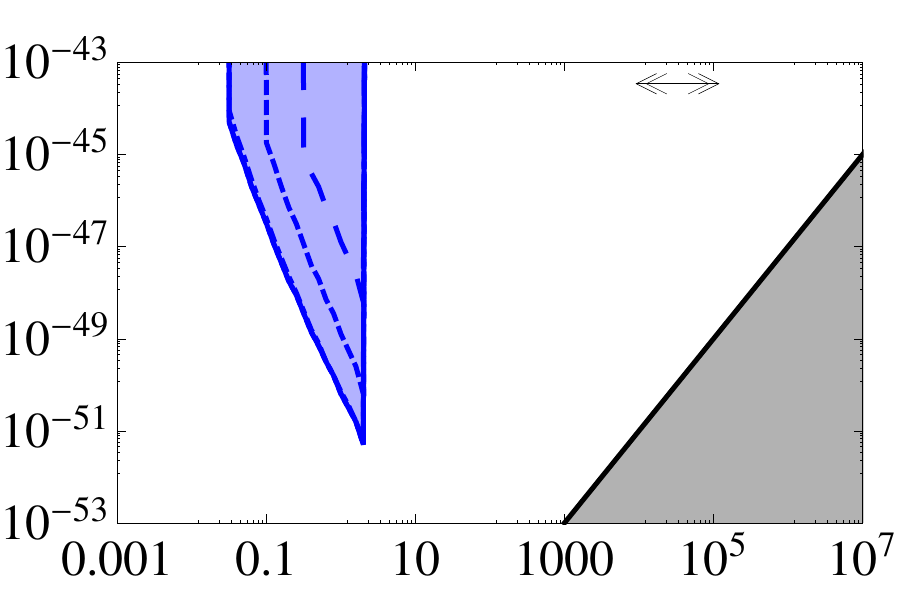} \end{minipage} &\begin{minipage}[c]{.4\textwidth}\includegraphics[scale=.7]{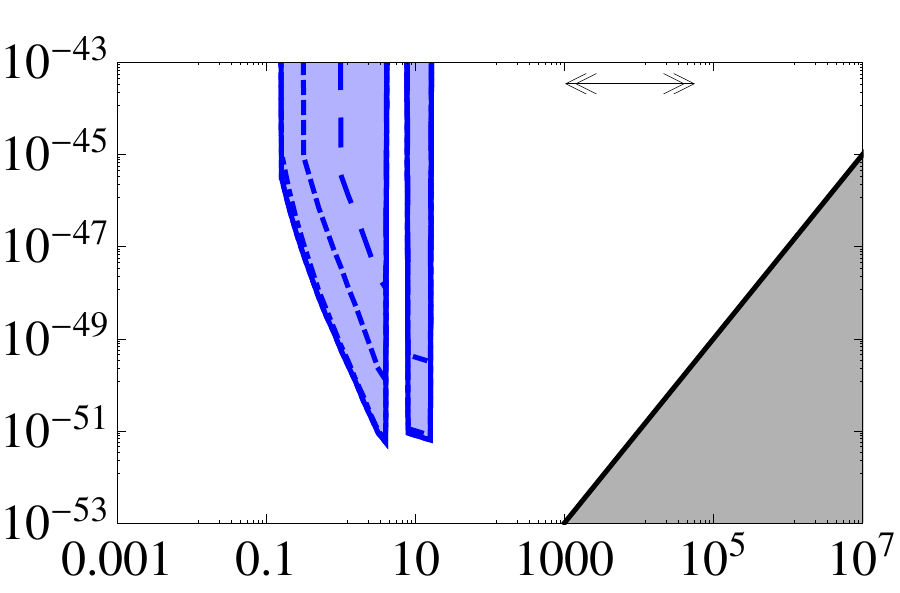} \end{minipage} &~~{\rotatebox[origin=c]{270}{$m_\phi = 1~{\rm MeV}$}}\\
&\begin{minipage}[c]{0.40\textwidth} \includegraphics[scale=.7]{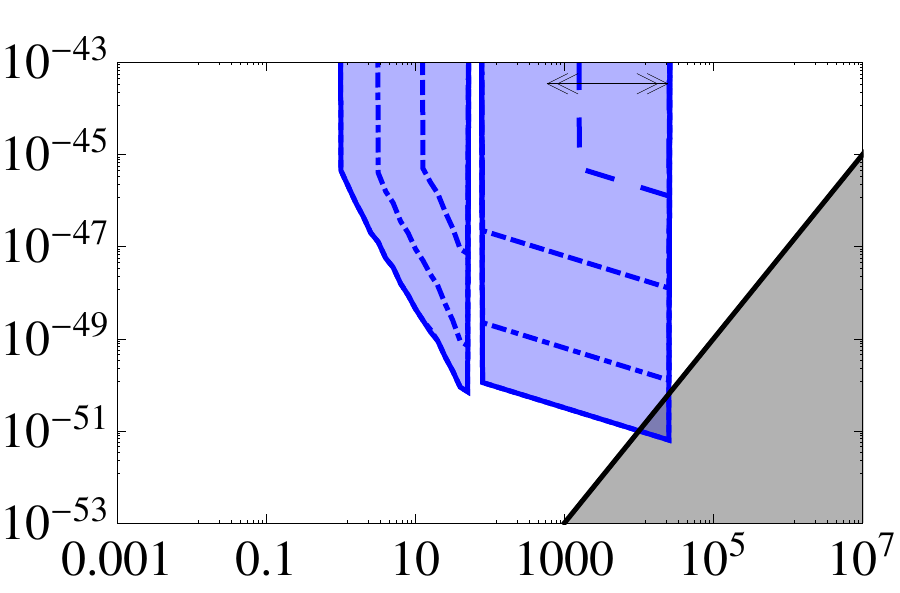} \end{minipage} &\begin{minipage}[c]{.4\textwidth}\includegraphics[scale=.7]{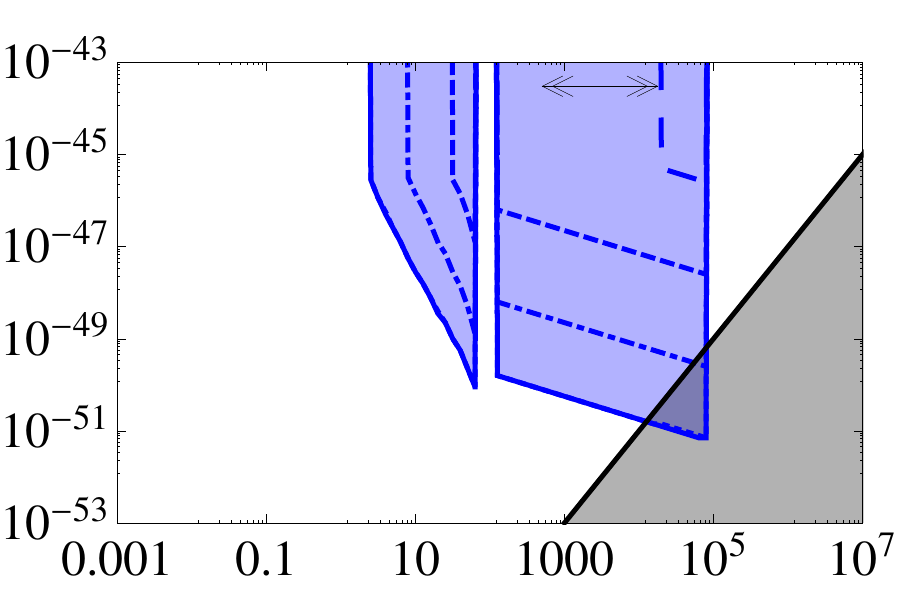} \end{minipage} &~~{\rotatebox[origin=c]{270}{$m_\phi = 5~{\rm MeV}$}}\\
&\begin{minipage}[c]{0.4\textwidth} \includegraphics[scale=.7]{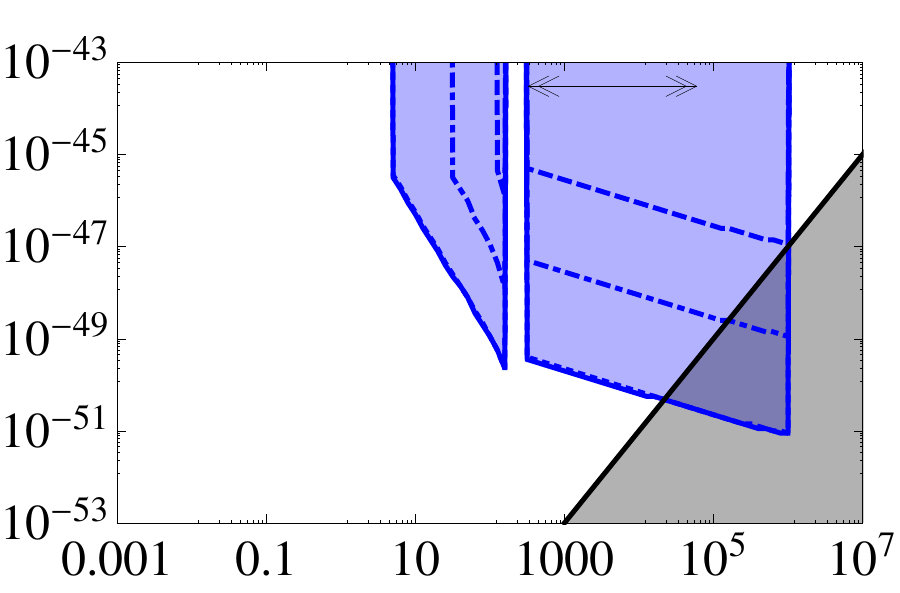} \end{minipage} &\begin{minipage}[c]{.4\textwidth}\includegraphics[scale=.7]{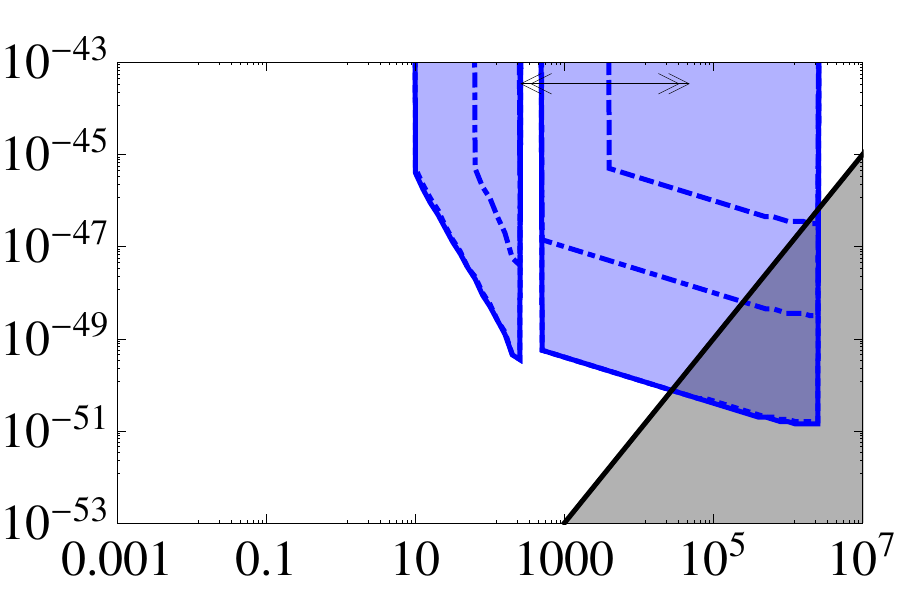} \end{minipage} &~~{\rotatebox[origin=c]{270}{$m_\phi = 10~{\rm MeV}$}}\\
 &  \multicolumn{2}{c}{Mass of Fermionic Dark Matter $m_X$ (GeV)} &
 \end{tabular}
\label{sigmavsmx}
\end{figure}

\begin{figure}
\caption{{\bf Co-annihilation:} Neutron star collapse bounds are shown in the $(m_X, \sigma_{nX})$ plane for dark matter fermions with an attractive Yukawa potential and a co-annihilation interaction with standard model baryons. These plots assume a 7 Gyr old neutron star with temperature $T=10^5 ~ \rm K$, in a local dark matter density $0.3~ \rm GeV / cm^3$ with average speed $\bar{v} = 220~ \rm km/s$. Constraint plots for Yukawa couplings $\alpha = \{10^{-1},10^{-2} \}$ and Yukawa boson masses $m_\phi = \{1,5,10  \}~{\rm MeV}$ are shown, and bounds on dark matter which co-annihilates with SM baryons at a rate $\Expect{\sigma_a v}_{co} = \{0,10^{-55},10^{-53}\} ~{\rm cm^3/s}$ are drawn with solid, dotted, and dotted-dashed lines, respectively. Horizontal arrows indicate the range of dark matter masses fitting dwarf galaxy mass halos (figure 6 in \cite{Tulin:2013teo}), assuming an attractive Yukawa coupling and the values of $\alpha$ and $m_\phi$ indicated. The shaded regions at the bottom of the plots indicate parameter space for which the bounds are lifted -- for heavier dark matter and a smaller scattering rate with nucleons, dark matter will thermalize at the center of the neutron star within 1 Gyr.}
 \begin{tabular}{cccc}
 & \begin{minipage}[c]{0.4\textwidth} \center ~~$\alpha =10^{-1}$ \end{minipage} &\begin{minipage}[c]{0.4\textwidth} \center ~~ $\alpha =10^{-2}$ \end{minipage} & \\
{\multirow{4}{*}{\rotatebox[origin=c]{90}{Nucleon-DM Scattering Cross Section $\sigma_{nX}$ $\rm (cm^2)$}}}
&\begin{minipage}[c]{0.40\textwidth} \includegraphics[scale=.70]{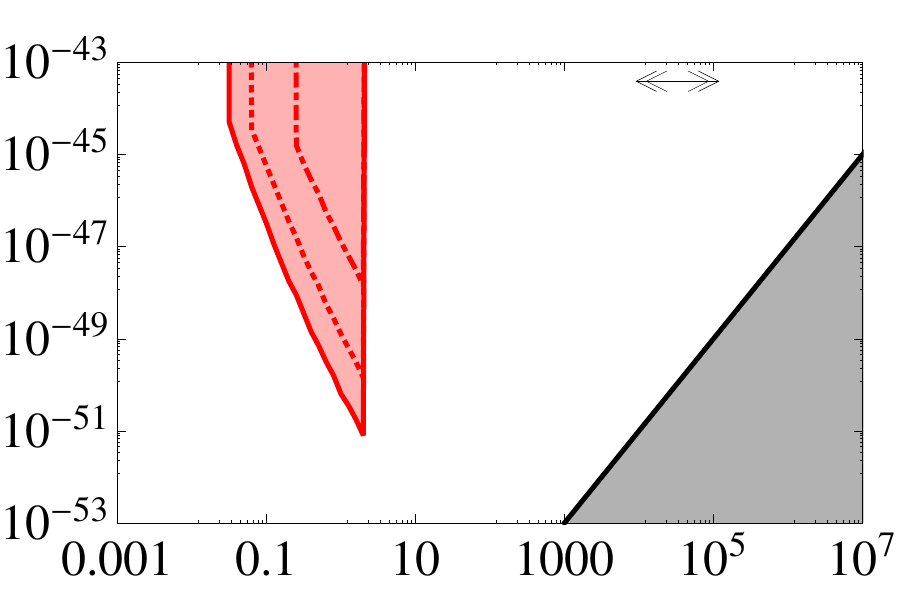} \end{minipage} &\begin{minipage}[c]{.4\textwidth}\includegraphics[scale=.7]{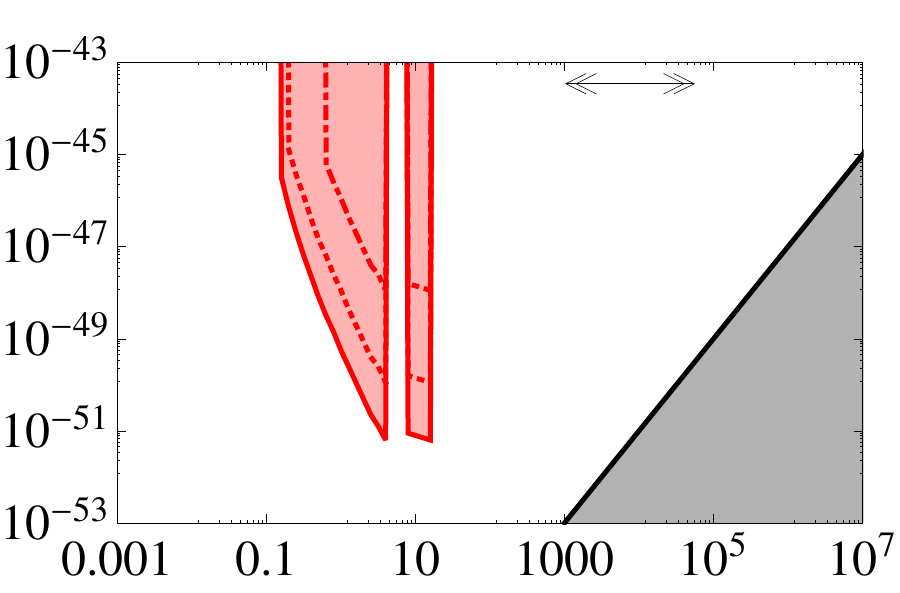} \end{minipage} &~~{\rotatebox[origin=c]{270}{$m_\phi = 1~{\rm MeV}$}}\\
&\begin{minipage}[c]{0.40\textwidth} \includegraphics[scale=.7]{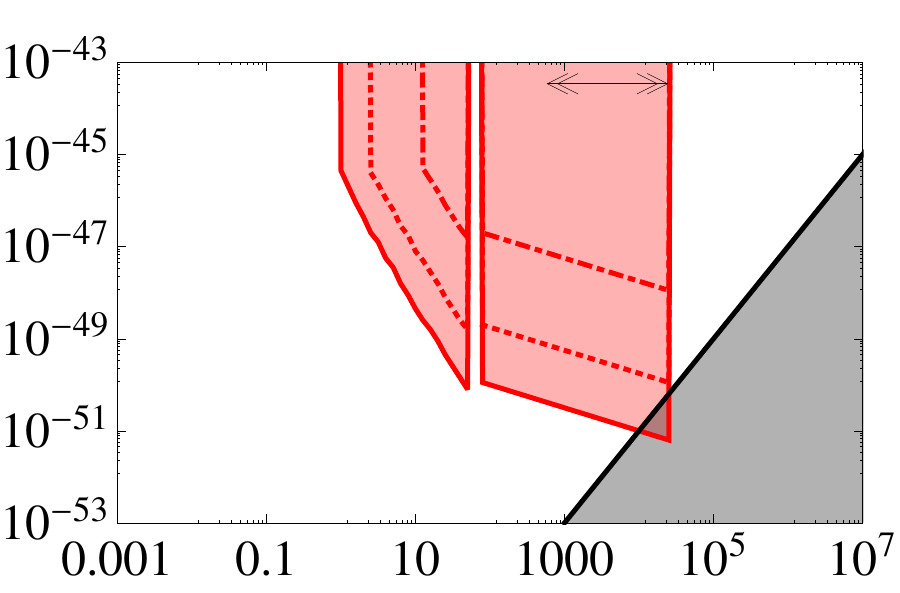} \end{minipage} &\begin{minipage}[c]{.4\textwidth}\includegraphics[scale=.7]{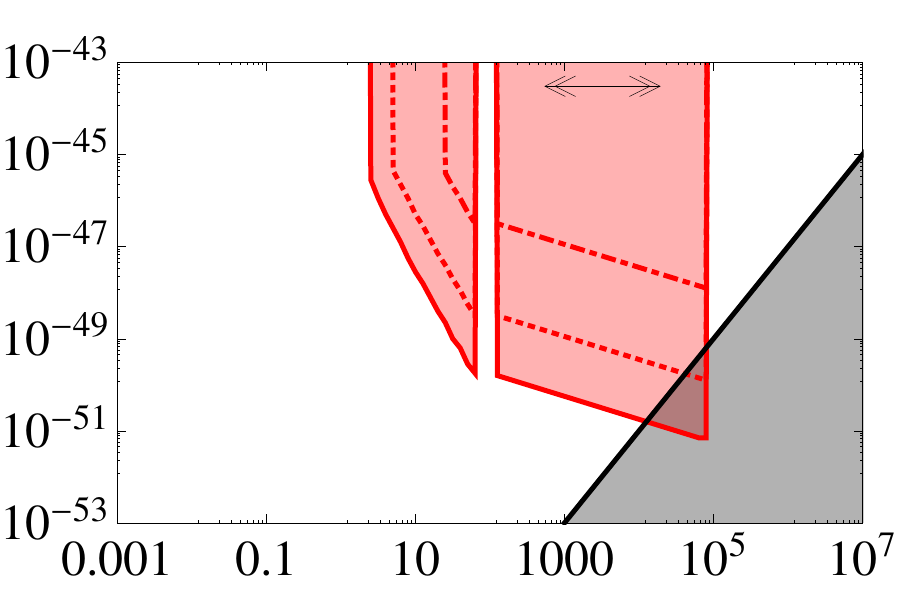} \end{minipage} &~~{\rotatebox[origin=c]{270}{$m_\phi = 5~{\rm MeV}$}}\\
&\begin{minipage}[c]{0.4\textwidth} \includegraphics[scale=.7]{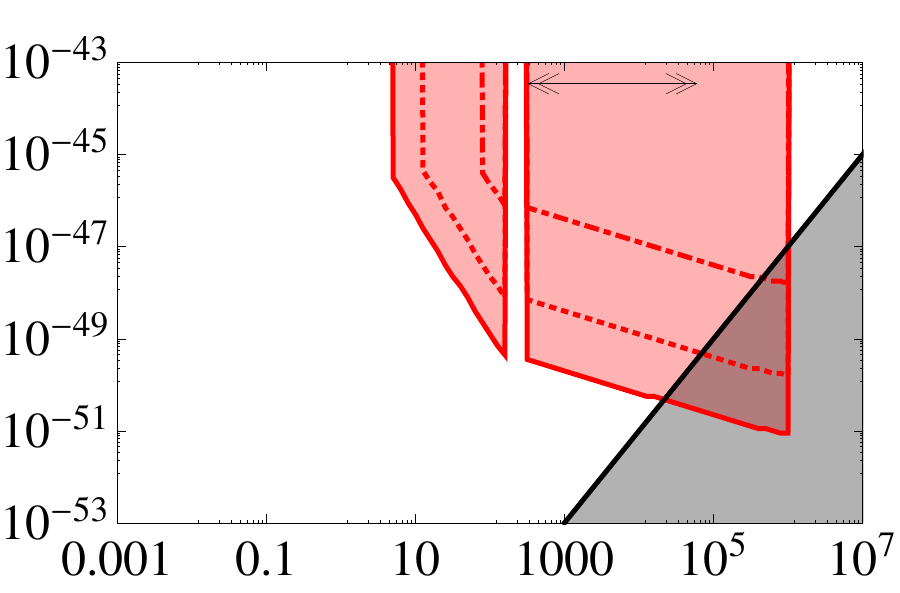} \end{minipage} &\begin{minipage}[c]{.4\textwidth}\includegraphics[scale=.7]{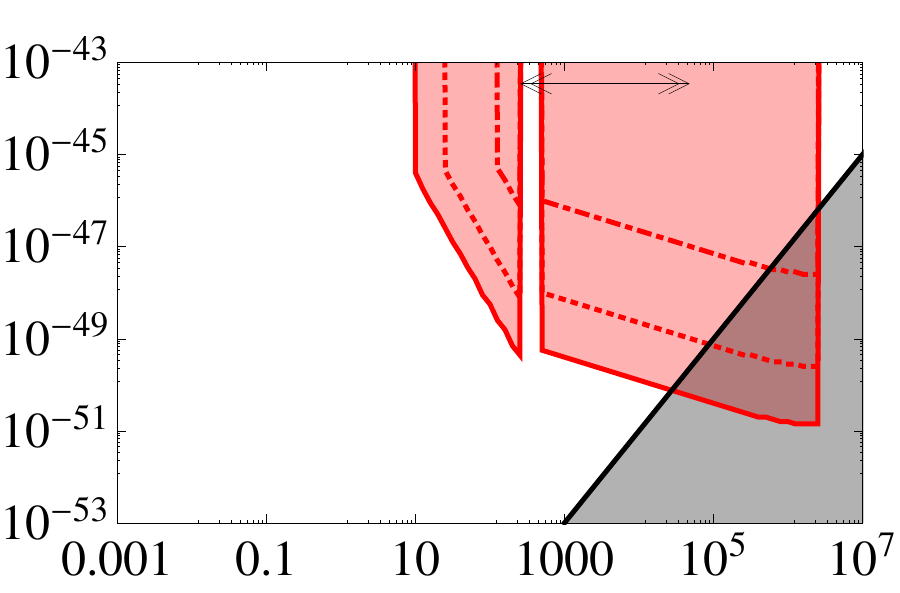} \end{minipage} &~~{\rotatebox[origin=c]{270}{$m_\phi = 10~{\rm MeV}$}}\\
 &  \multicolumn{2}{c}{Mass of Fermionic Dark Matter $m_X$ (GeV)} &
 \end{tabular}
\label{sigmavsmxcoann}
\end{figure}

Figure \ref{sigmavsmx} shows exclusion contours for dark matter fermions with an attractive Yukawa coupling in the ($m_X, \sigma_{nX}$) plane, assuming Yukawa mediator masses $m_\phi = \{1,5,10 \} ~ \rm MeV$ and $\alpha = \{10^{-1},10^{-2} \}$.  A local dark matter density of $\rho_X = 0.3 \rm ~GeV/cm^3$ with average speed $\bar{v} = 220 \rm ~ km/s$ is assumed along with a neutron star 7 Gyr old, with temperature $10^5$ K, density $\rho_b = 10^{38} \rm GeV/cm^3 $, and radius $R_n = 10 \rm ~ km$. It is evident in Figure \ref{sigmavsmx} that a stronger Yukawa coupling constant $\alpha$ shifts bounds to smaller $m_X$, while a larger Yukawa mediator mass $m_\phi$ shifts and broadens bounds over larger $m_X$.
This behavior is easy to understand.  The dark matter capture rate scales roughly as $\propto m_X^{-1}$, while for the relevant regions of parameter
space, the number of dark matter particles needed for collapse scales with a more negative power of $m_X$.  As a result, for smaller $m_X$,
collapse can only occur if there is an enhancement in the Yukawa force.
Bounds on $\sigma_{nX}$ also become tighter as $m_X$ increases, because less dark matter is required for collapse.
The mass of the black hole which is formed by collapse thus decreases as $m_X$ increases.
Note that for the fixed choices of $\alpha$ and $m_\phi$ in
figure~\ref{sigmavsmx}, bounds on $\sigma_{nX}$ cut off sharply at large $m_X$.  This occurs because for large enough $m_X$, the mass of the black hole
produced is so small that it evaporates quickly and does not consume the neutron star.

The solid, dotted, dotted-dashed, short-dashed, and long-dashed curves show the parameter space which would be excluded if the
dark matter had a self-annihilation cross section of $\Expect{\sigma_a v} = \{0,10^{-50},10^{-47},10^{-45},10^{-43}\} ~{\rm cm^3/s}$.
These curves make it apparent that only a modest amount of dark matter self-annihilation is required to eliminate neutron star bounds on
fermionic self-interacting dark matter. In fact, all bounds shown in this paper are
entirely eliminated if $\Expect{\sigma_a v} = 10^{-40} \rm ~ cm^3/s$.  Of course this lifting of the
bound assumes that the dark matter annihilates either to standard model particles or some dark sector radiation which does not participate in dark matter collapse.

In Figure \ref{sigmavsmxcoann} we consider the same exclusion contours, but for dark matter fermions which co-annihilate with nucleons. We assume the same neutron star parameters (7 Gyr old, $10^5$ K, 0.3 $\rm GeV/cm^3$ ambient dark matter density with average velocity $220~\rm km/s$) In this case the co-annihilation cross sections required to lift the bounds are much smaller -- we plot $\Expect{\sigma_a v}_{co} =  \{0,10^{-55},10^{-53}\} ~{\rm cm^3/s}$ with solid, dotted, and dotted-dashed lines respectively. A co-annihilation cross section of $\Expect{\sigma_a v}_{co} = 10^{-50} ~{\rm cm^3/s}$ is all that is needed to eliminate every bound shown in this paper.

For both self- and co-annihilation, we find that assuming an old neutron star sits in a much denser background of dark matter, for instance $\rho_X = 10^3 \rm ~ GeV/cm^3$ as is
 possible for the core of a globular cluster, the bounds on the cross section will increase proportionally by a factor of $\sim 10^3$.
 The net result is that even for an old neutron star bombarded by globular cluster densities of dark matter, the required self- and co-annihilations necessary to eliminate
all constraints are still tiny: $\Expect{\sigma_a v} = 10^{-37} \rm ~ cm^3/s$ and $\Expect{\sigma_a v}_{co} \sim 10^{-47} ~{\rm cm^3/s}$  respectively.

These bounds on Yukawa-coupled dark matter extend over intermediate masses not excluded in prior work \cite{Kouvaris:2011gb}. As explained in section \ref{sectioncollapse}, considering collapse from a strongly-screened non-degenerate state through a strongly-screened degenerate state excludes this dark matter phase space.\footnote{This corresponds to the rightmost collapse bound channel shown in the collapse bound table.} For some of the low-mass dark matter exclusion contours, cross sections larger than $10^{-45}~ \rm cm^2$ are bounded, which is nominally at odds with a hitherto common lore that the maximum geometric cross section for dark matter traveling through a neutron star is $10^{-45}~ \rm cm^2$. This is not true for low-mass dark matter particles which interact with fewer neutrons along their path due to Pauli-blocking \cite{Bell:2013xk}, as we discuss in Section \ref{secacc}.

In these plots as $m_X$ increases, the state of the dark matter at the beginning of collapse changes from a partly-screened degenerate state,
to a strongly-screened degenerate state, to a non-degenerate state collapsing through a strongly-screened degenerate state, and finally
to a non-degenerate state collapsing through a partly-screened degenerate state.
Dark matter can only collapse from a degenerate state if a large number of dark matter particles have been captured, implying that $m_X$
will be smaller than for non-degenerate collapse.  Moreover, the dark matter can only be partly-screened if the
interparticle spacing is small, implying that the amount of captured dark matter is large, and in turn that $m_X$ is small. For the dark matter and scalar mediator masses we consider, bounds only arise after collapse from a partly-screened, degenerate phase when $m_\phi < 10 ~\rm{MeV}$ and $\alpha < 10^{-3}$, as in the leftmost bound regions of the top two panels of Figures \ref{fig4} and \ref{fig7}.

In most of the plots there are apparent gaps between low and high mass bounds; these low and high mass bounds correspond to collapse initiating from degenerate and non-degenerate phases, respectively. The gaps are regions of parameter space where collapse initiates for strongly-screened non-degenerate dark matter, but then condition \eqref{eq2nduscoll} is not met. We nevertheless expect that old neutron stars bound these gaps in parameter space, and that the gaps are an artifact of treating the dark matter as either entirely degenerate or entirely non-degenerate. Future work could explicitly bound this parameter space by considering partly-degenerate dark matter and simulating the full dynamics of collapse from a strongly-screened non-degenerate phase. 

It is worth noting that Figures \ref{sigmavsmx} and \ref{sigmavsmxcoann} constrain attractively self-interacting dark matter tuned
to fit flat-cored dwarf galaxies~\cite{Tulin:2012wi,Tulin:2013teo}. In the figures we indicate the relevant dark matter mass
range from \cite{Tulin:2013teo} with fletched arrows. For a Yukawa coupling of $0.1-0.01$, the models favor $m_\phi \sim 1-10$ MeV with
a dark matter mass of $m_X = 10-10^5~{\rm GeV}$. As can be seen in the aforementioned panels, models of attractively Yukawa-coupled dark matter fermions
in the bound region must either have an undetectably small scattering cross section with Standard Model particles or have a small annihilation interaction
to deplete the amount of dark matter in the neutron star.

Dark matter capture in stars can be boosted by self-scattering off already captured DM (see e.g. \cite{Zentner:2009is}). The recent work of \cite{Guver:2012ba} shows that this affect can substantially increase the number of captured dark matter particles for a large self-interaction capture cross-section ($\sigma_{XX} = 10^{-24}~ \rm{cm^2}$), if one also assumes a smaller dark matter-nucleon cross section than those bounded in this work ($\sigma_{nX} \lesssim 10^{-55}~ \rm{cm^2}$). While the light scalar mediators considered in this work imply self-scattering cross-sections as large as $\sigma_{XX} = 10^{-24} ~\rm{cm^2}$ at low velocities, dark matter falling into a neutron star will be relativistic and therefore have a self-scattering cross-section much smaller than $10^{-24} ~\rm{cm^2}$, as detailed in \cite{Tulin:2013teo}. We leave a thorough analysis of stellar dark matter self-capture for light mediator models to future work.

In Appendix \ref{appfurtherbounds} we expand the bounds plotted to parameter space with $\alpha = \{10^{-3},10^{-4}\}$ and $m_\phi = \{50, 100, 500 ~ \rm MeV \}$.

\section{Conclusions} \label{secconclusion}

We have derived neutron star bounds on fermion dark matter which scatters with nucleons and self-interacts via an attractive  Yukawa coupling. The plots in Section \ref{sectionconstraints} focus on parameter space favored by dark matter models that utilize a Yukawa-mediated self-interaction to explain the cored profiles of dwarf galaxies and provide an appropriate level of freeze-out annihilation during reheating \cite{Tulin:2013teo}. A key feature of these models is that the dark matter self-interaction cross section will have a velocity-dependence set by the mass of the dark matter and Yukawa mediator, $\phi$, leading to different dynamics for slower dwarf galaxy halos and faster galactic cluster halos. Indeed, such models can even account for DM-neutron scattering if one assumes at least one new state with both Yukawa mediator $\phi$ charge and a standard model gauge charge \cite{Kaplinghat:2013kqa}. Although a full analysis of a fermion dark sector observable in all detection channels (direct, indirect, collider), with a simple characterization -- a Yukawa coupling and Yukawa charge mixing with SM charge -- is beyond the aims of this paper, such a model would be constrained by observations of old neutron stars.

Besides bounding dark matter tuned to explain dwarf galaxy mass profiles, neutron stars also bound generic fermion dark matter with an attractive Yukawa coupling $10^{-1}-10^{-4}$ and a mediator mass $1-500 \rm~ MeV$ over a broad range of dark matter masses $0.01-10^7 \rm ~ GeV$ as elaborated in Section \ref{sectionconstraints} and Appendix \ref{appfurtherbounds}. Nevertheless, in this paper we have shown the rather simple condition required to lift bounds on fermion dark matter: a self-annihilation interaction of $10^{-40} \rm ~cm^3/s$ or a neutron co-annihilation cross section of $10^{-50} \rm ~cm^3/s$.

Note that the self-annihilation cross section needed to completely eliminate bounds on fermion dark matter is about an order of magnitude
larger than that required to eliminate all bounds on bosonic dark matter.  This reflects the fact that a sufficiently strong attractive
Yukawa interaction can more than overcome the Fermi degeneracy pressure.  Nevertheless, the magnitude of the self-annihilation and
co-annihilation cross sections needed to eliminate bounds from neutron star observations is so small that the annihilation or co-annihilation
processes are unlikely to be observed at any current or near-future indirect detection experiments.

If attractive self-interactions are strong enough, then a pair of fermion dark matter particles can potentially form an
integer-spin bound state.  This process may have an interesting effect on dark matter bounds from observations of old
neutron stars, which we leave as an topic for future study.

\vskip .2in
\textbf{Acknowledgments}

We are grateful to Jessica Cook, Jon Sapirstein, Tim Tait, Xerxes Tata and Haibo Yu for useful discussions. The work of J.~B.~is supported in part by a UND Theoretical High Energy Physics Fellowship. The work of J.~K.~is supported in part by NSF CAREER Award PHY-1250573.

\appendix

\section{Dynamics for a screened Yukawa potential} \label{appyukawa}

In this appendix we establish a new variable $y$ (as in \cite{Kouvaris:2011gb}) and reframe the virial equation for fermion dark matter at a neutron star's core in the limits $y>1$ and $y<1$. First we identify the nearest-neighbor inter-particle distance for $N_X$ dark matter particles evenly spaced in a sphere of radius $r$: $x \equiv r \left(\frac{4\pi}{3N}\right)^{1/3} \simeq 1.6 r N_X^{-1/3}$ is the nearest-neighbor inter-particle distance. Then, in order to recast Eq.~\eqref{virialfull} into an equation dependent only on $r$ and not on sums over inter-particle separation, $r_j$, we consider the argument of the exponential terms ($m_\phi r_j$) in two limits.
\begin{itemize}
\item[1.] ${\bf m_\phi r_j \gtrsim 1}$ \textit{Strongly-screened Yukawa potential.} If for all inter-particle distances, the Yukawa potential is strongly-screened by the exponential term, the strength of the Yukawa potential will depend mostly on nearest neighbor dark matter fermions. Here we define a dimensionless variable, the shortest inter-particle distance multiplied by the mediator mass, $y \equiv 1.6 m_\phi r / N^{1/3}$.
    Assuming the fermions are evenly spaced, with $\sim 8$ nearest neighbors, the Yukawa potential can be approximated as $V_{\rm Yuk.,scr}
    \simeq -8 \alpha m_\phi e^{-y}/y $. With this Yukawa potential and making the substitution ($r \rightarrow \frac{y N_X^{1/3}}{(4/3 \pi)^{1/3}
    m_\phi}$) Eq.~\eqref{virialfull} becomes
\begin{align}
-2 E_k + \frac{(4/3 \pi)^{1/3} G N_X^{2/3} \rho_b  m_X y^2}{m_\phi^2} + \frac{(4/3 \pi)^{1/3} G N_X^{2/3} m_X^2 m_\phi}{y}  + 8\alpha \left( \frac{m_\phi e^{-y}}{y} + m_\phi e^{-y} \right)=0 \label{virialscreenedapp}.
\end{align}
\item[2.] ${\bf m_\phi r_j \lesssim  1}$ \textit{Partly-screened Yukawa potential.} In phase space where the inter-particle distances $r_j$ are smaller than the inverse of the Yukawa mediator mass, $r_j < 1/m_\phi$, it is safe to approximate the partly-screened Yukawa potential as a Coulomb potential because the exponential term does not appreciably screen the particles inside the radius ``$1/m_\phi$". The number of these unscreened particles can be estimated; noting that the inverse of the shortest inter-particle distance $1/x \simeq N_X^{1/3}/(1.6 r)$ is the linear density of dark matter particles in the center of the neutron star, the number of unscreened particles is $N_{us} \simeq \frac{4}{3} \pi/(m_\phi x)^3 = \frac{4}{3} \pi/y^3$. The partly-screened Yukawa potential resembles a Coulomb potential with radius $1/m_\phi$~\cite{Kouvaris:2011gb},
\begin{align}
V_{\rm Yuk.,ps} \simeq -\frac{N_{us}\alpha}{\frac{1}{m_\phi}}=-\frac{\alpha N_X}{m_\phi^2 r^3} = -\frac{4 \pi \alpha m_\phi}{3 y^3},
\end{align}
and the virial equation for the partly-screened Yukawa potential is
\begin{align}
-2 E_k + \frac{(4/3 \pi)^{1/3} G N_X^{2/3} \rho_b  m_X y^2}{m_\phi^2} + \frac{(4/3 \pi)^{1/3} G N_X^{2/3} m_X^2 m_\phi}{y}  + \frac{4 \pi \alpha m_\phi}{y^3}=0 \label{virialpsapp}.
\end{align}
\end{itemize}

\section{Further neutron star dark matter bound plots}\label{appfurtherbounds}
This appendix displays bounds on self-annihilating and co-annihilating fermionic dark matter with an attractive Yukawa self-interaction in the ($m_X, \sigma_{nX}$) plane for the following slices of parameter space, $\alpha = \{10^{-1},10^{-2},10^{-3},10^{-4} \}$, $m_\phi = \{1,5,10,50,100,500 \} \rm ~ MeV$, and either a self-annihilation of $\Expect{\sigma_a v} = \{0,10^{-50},10^{-47},10^{-45},10^{-43} \}~{\rm cm^3/s}$, shown as blue solid, dotted, dotted-dashed, short dashed, and long dashed lines, or a co-annihilation of $\Expect{\sigma_a v}_{co} = \{0,10^{-55},10^{-53}\}~{\rm cm^3/s}$ shown as red solid, dotted, and dotted-dashed lines.

\begin{figure}
\caption{{\bf Self-annihilation:} Neutron star collapse bounds are shown in the $(m_X, \sigma_{nX})$ plane for self-annihilating dark matter fermions with an attractive Yukawa-coupling . Yukawa couplings $\alpha = \{10^{-1},10^{-2} \}$ and Yukawa boson masses $m_\phi = \{50,100,500  \}~{\rm MeV}$ are displayed, and bounds on dark matter with a self-annihilation cross section of $\Expect{\sigma_a v} = \{10^{0,-50},10^{-47},10^{-45},10^{-43}\} ~{\rm cm^3/s}$ are drawn with solid, dotted, dotted-dashed, short-dashed, and long-dashed lines, respectively. The shaded grey regions at the bottom right of each plot are unbounded because the dark matter
has insufficient time to thermalize, as detailed in Section \ref{sectioncollapse}.}
 \begin{tabular}{cccc}
 & \begin{minipage}[c]{0.4\textwidth} \center ~~$\alpha =10^{-1}$ \end{minipage} &\begin{minipage}[c]{0.4\textwidth} \center ~~ $\alpha =10^{-2}$ \end{minipage} & \\
{\multirow{4}{*}{\rotatebox[origin=c]{90}{Nucleon-DM Scattering Cross Section $\sigma_{nX}$ $\rm (cm^2)$}}}
&\begin{minipage}[c]{0.40\textwidth} \includegraphics[scale=.70]{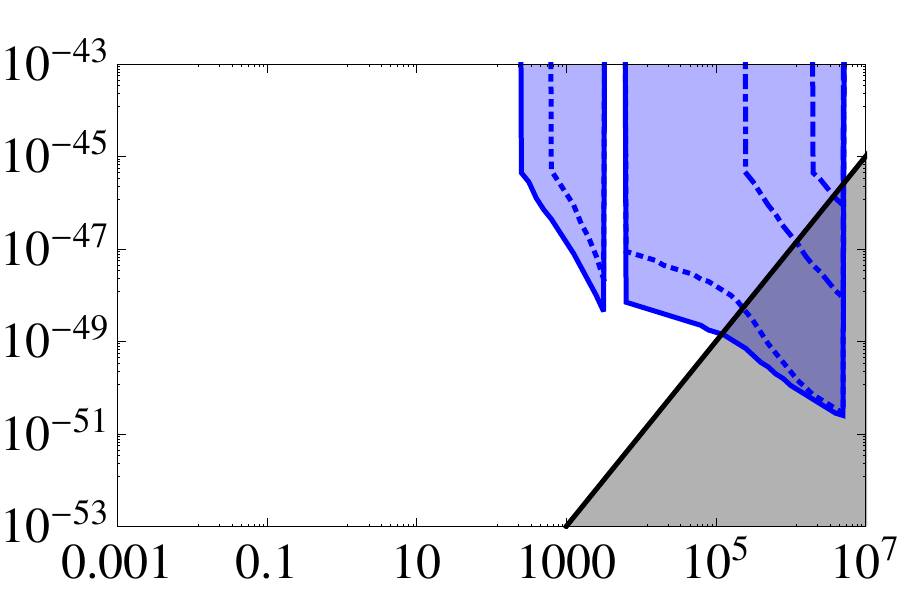} \end{minipage} &\begin{minipage}[c]{.4\textwidth}\includegraphics[scale=.70]{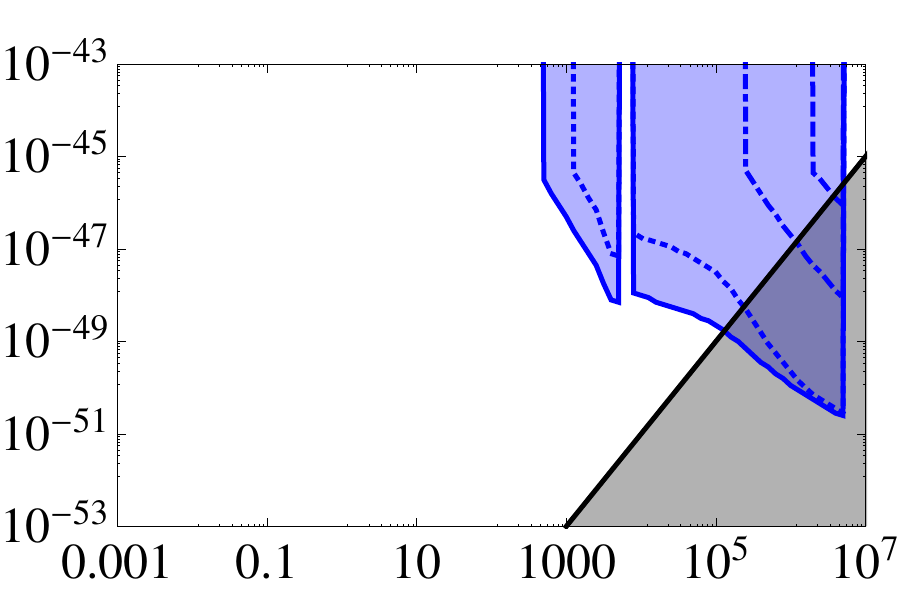} \end{minipage} &~~{\rotatebox[origin=c]{270}{$m_\phi = 50~{\rm MeV}$}}\\
&\begin{minipage}[c]{0.40\textwidth} \includegraphics[scale=.7]{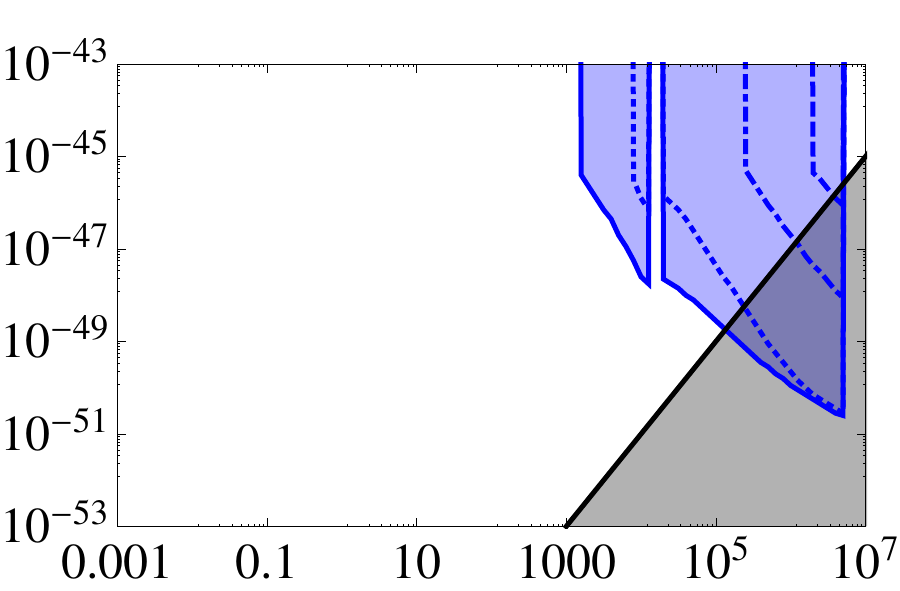} \end{minipage} &\begin{minipage}[c]{.4\textwidth}\includegraphics[scale=.7]{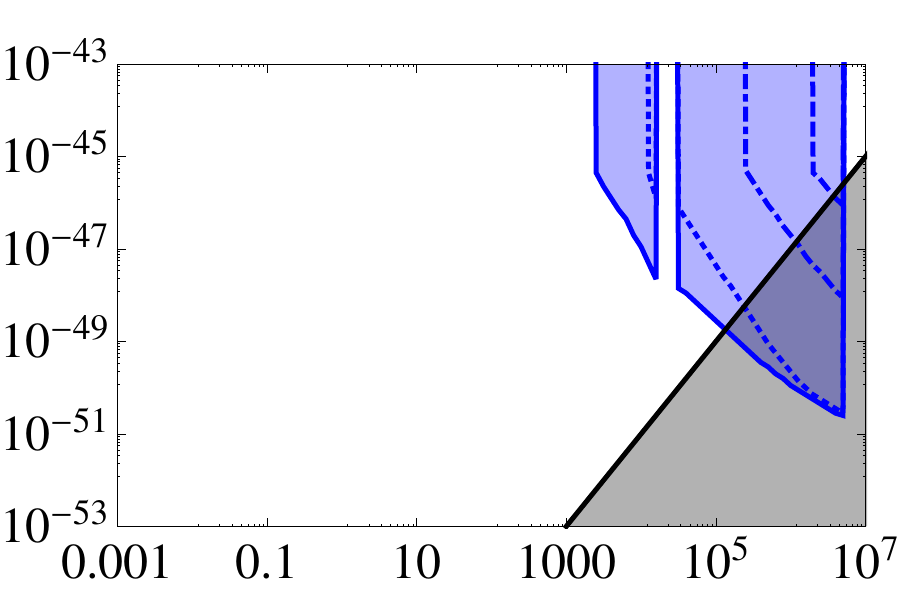} \end{minipage} &~~{\rotatebox[origin=c]{270}{$m_\phi = 100~{\rm MeV}$}}\\
&\begin{minipage}[c]{0.4\textwidth} \includegraphics[scale=.7]{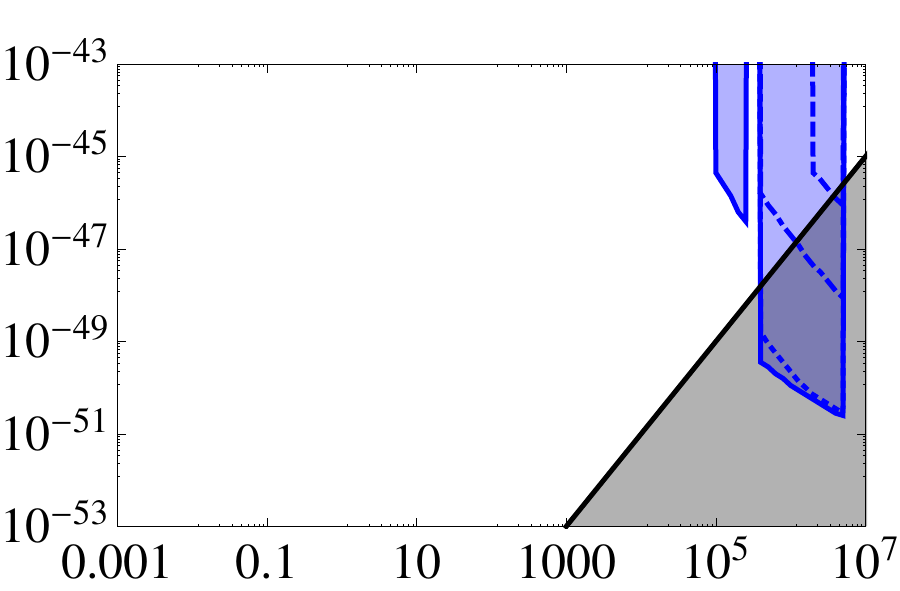} \end{minipage} &\begin{minipage}[c]{.4\textwidth}\includegraphics[scale=.7]{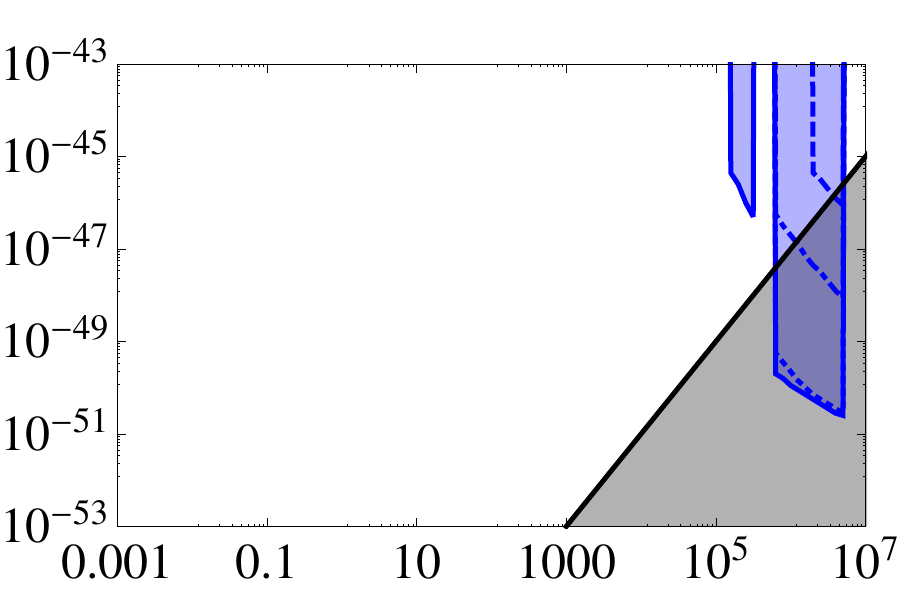} \end{minipage} &~~{\rotatebox[origin=c]{270}{$m_\phi = 500~{\rm MeV}$}}\\
 &  \multicolumn{2}{c}{Mass of Fermionic Dark Matter $m_X$ (GeV)} &
 \end{tabular}
\end{figure}

\begin{figure}
\caption{{\bf Self-annihilation:} Neutron star collapse bounds are shown in the $(m_X, \sigma_{nX})$ plane for self-annihilating dark matter fermions with an attractive Yukawa-coupling . Yukawa couplings $\alpha = \{10^{-3},10^{-4} \}$ and Yukawa boson masses $m_\phi = \{1,5,10  \}~{\rm MeV}$ are displayed, and bounds on dark matter with a self-annihilation cross section of $\Expect{\sigma_a v} = \{0,10^{-50},10^{-47},10^{-45},10^{-43}\} ~{\rm cm^3/s}$ are drawn with solid, dotted, dotted-dashed, short-dashed, and long-dashed lines, respectively. The shaded grey regions at the bottom right of each plot are unbounded because the dark matter has
insufficient time to thermalize, as detailed in Section \ref{sectioncollapse}.}
 \begin{tabular}{cccc}
 & \begin{minipage}[c]{0.4\textwidth} \center ~~$\alpha =10^{-3}$ \end{minipage} &\begin{minipage}[c]{0.4\textwidth} \center ~~ $\alpha =10^{-4}$ \end{minipage} & \\
{\multirow{4}{*}{\rotatebox[origin=c]{90}{Nucleon-DM Scattering Cross Section $\sigma_{nX}$ $\rm (cm^2)$}}}
&\begin{minipage}[c]{0.40\textwidth} \includegraphics[scale=.70]{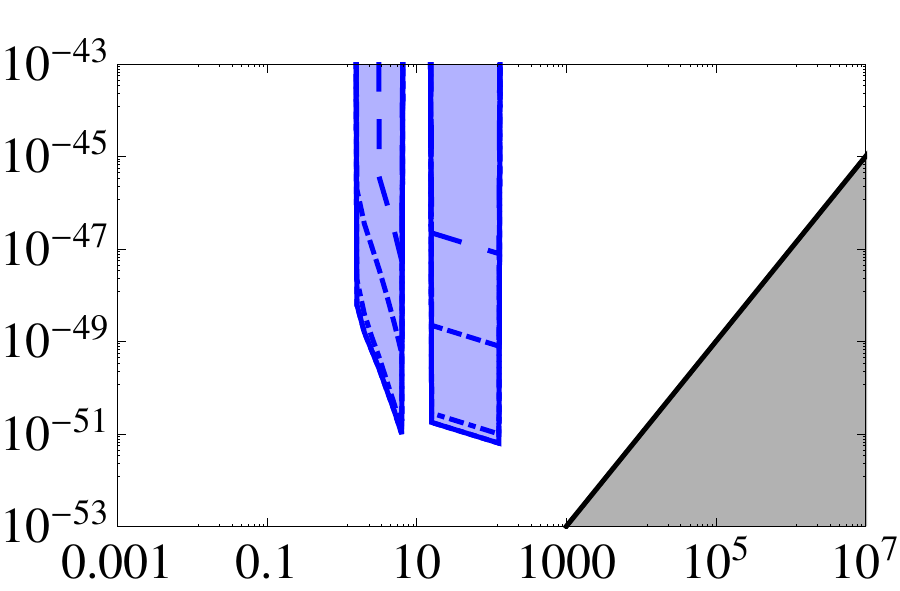} \end{minipage} &\begin{minipage}[c]{.4\textwidth}\includegraphics[scale=.70]{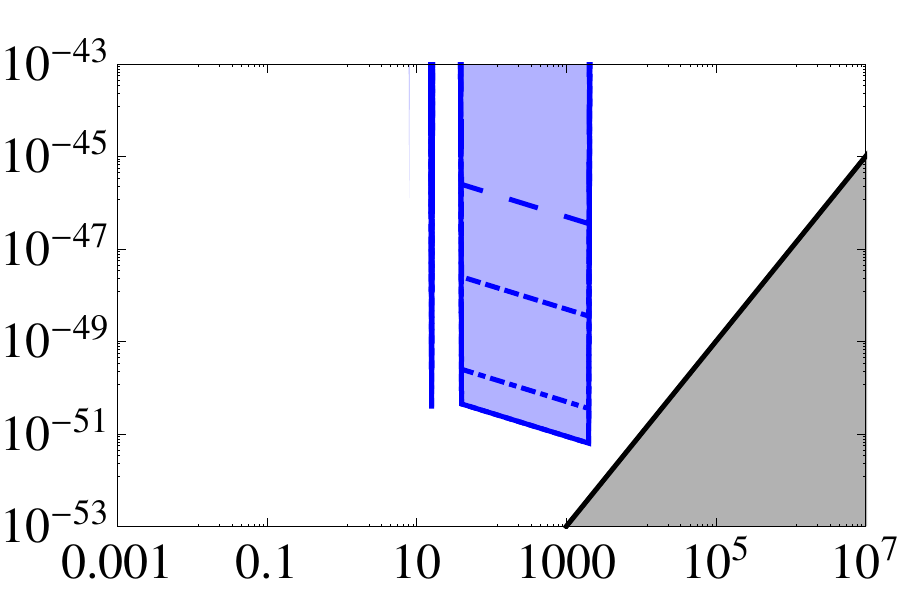} \end{minipage} &~~{\rotatebox[origin=c]{270}{$m_\phi = 1~{\rm MeV}$}}\\
&\begin{minipage}[c]{0.40\textwidth} \includegraphics[scale=.7]{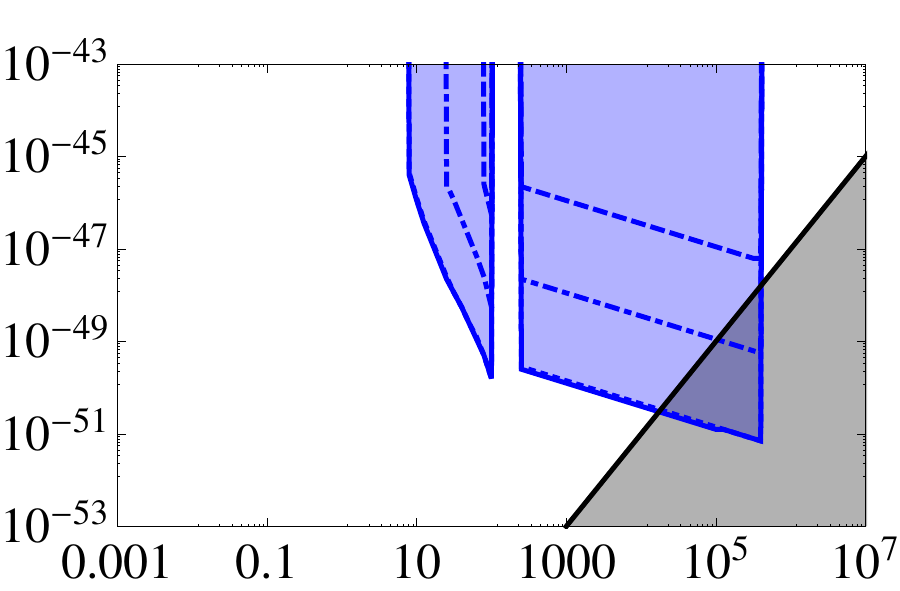} \end{minipage} &\begin{minipage}[c]{.4\textwidth}\includegraphics[scale=.7]{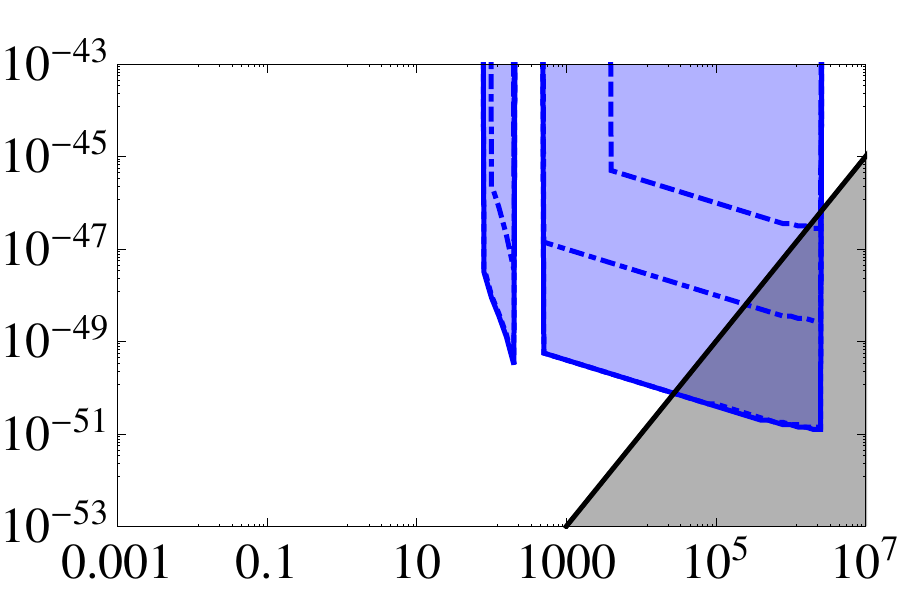} \end{minipage} &~~{\rotatebox[origin=c]{270}{$m_\phi = 5~{\rm MeV}$}}\\
&\begin{minipage}[c]{0.4\textwidth} \includegraphics[scale=.7]{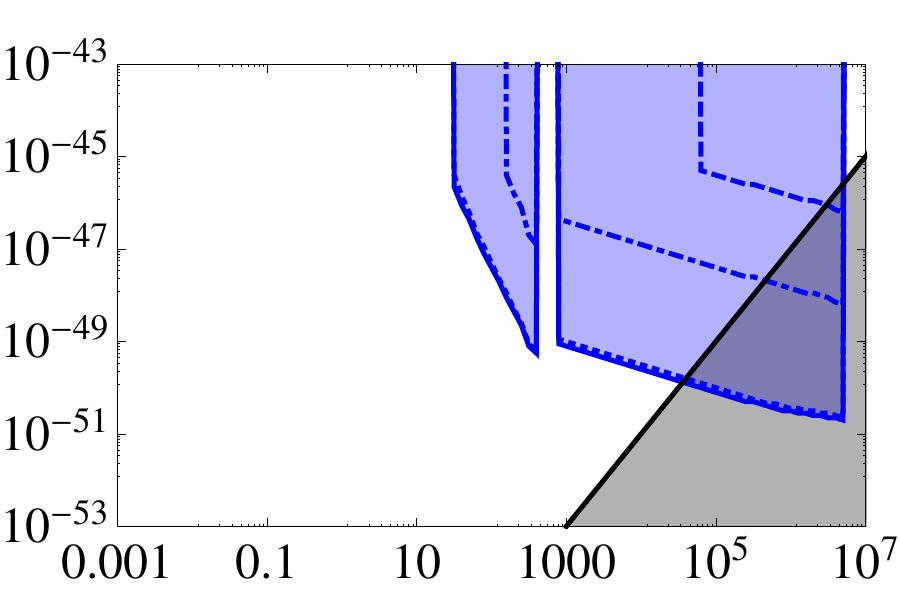} \end{minipage} &\begin{minipage}[c]{.4\textwidth}\includegraphics[scale=.7]{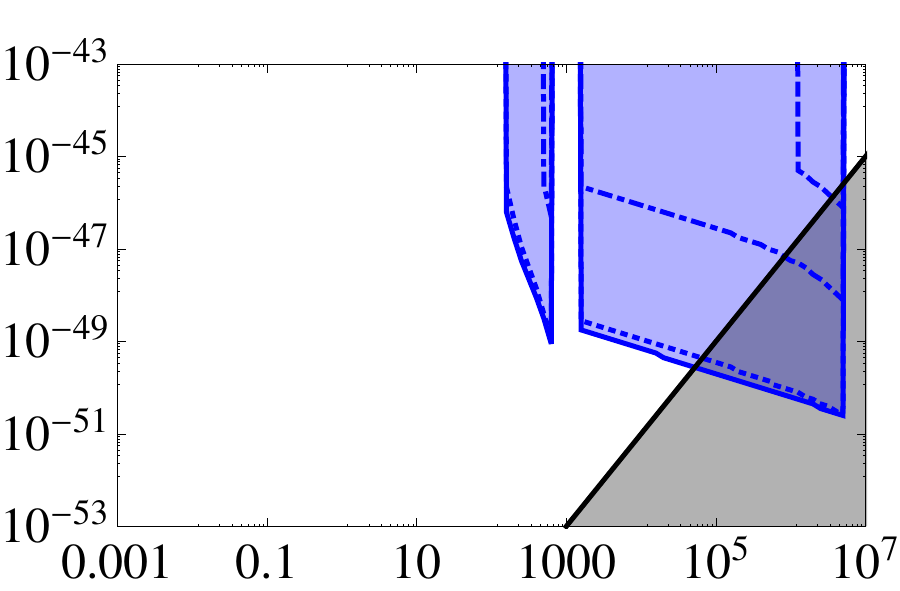} \end{minipage} &~~{\rotatebox[origin=c]{270}{$m_\phi = 10~{\rm MeV}$}}\\
 &  \multicolumn{2}{c}{Mass of Fermionic Dark Matter $m_X$ (GeV)} &
 \end{tabular}
 \label{fig4}
\end{figure}

\begin{figure}
\caption{{\bf Self-annihilation:} Neutron star collapse bounds are shown in the $(m_X, \sigma_{nX})$ plane for self-annihilating dark matter fermions with an attractive Yukawa-coupling . Yukawa couplings $\alpha = \{10^{-3},10^{-4} \}$ and Yukawa boson masses $m_\phi = \{50,100,500  \}~{\rm MeV}$ are displayed, and bounds on dark matter with a self-annihilation cross section of $\Expect{\sigma_a v} = \{0,10^{-50},10^{-47},10^{-45},10^{-43}\} ~{\rm cm^3/s}$ are drawn with solid, dotted, dotted-dashed, short-dashed, and long-dashed lines, respectively. The shaded grey regions at the bottom right of each plot are unbounded because the dark matter has
insufficient time to thermalize, as detailed in Section \ref{sectioncollapse}.}
 \begin{tabular}{cccc}
 & \begin{minipage}[c]{0.4\textwidth} \center ~~$\alpha =10^{-3}$ \end{minipage} &\begin{minipage}[c]{0.4\textwidth} \center ~~ $\alpha =10^{-4}$ \end{minipage} & \\
{\multirow{4}{*}{\rotatebox[origin=c]{90}{Nucleon-DM Scattering Cross Section $\sigma_{nX}$ $\rm (cm^2)$}}}
&\begin{minipage}[c]{0.40\textwidth} \includegraphics[scale=.70]{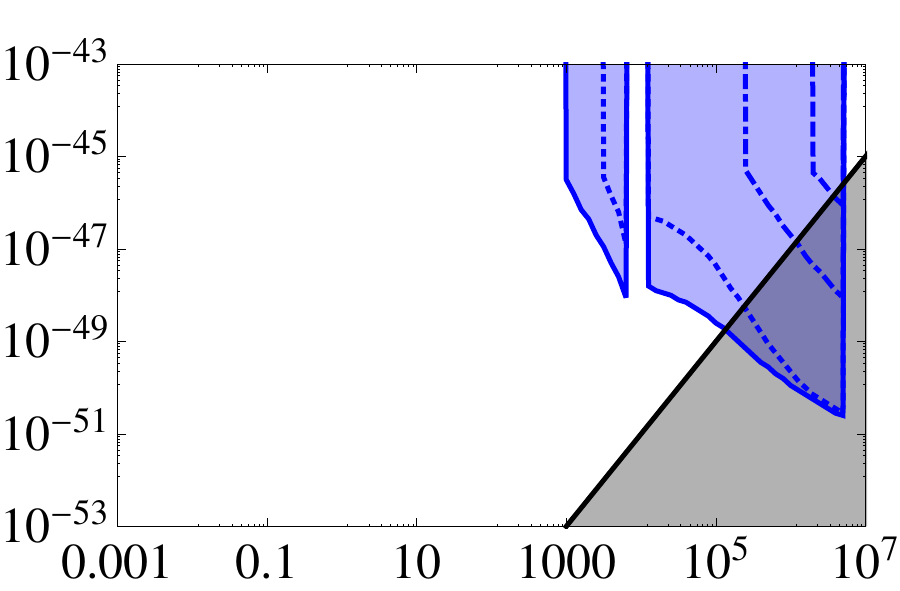} \end{minipage} &\begin{minipage}[c]{.4\textwidth}\includegraphics[scale=.70]{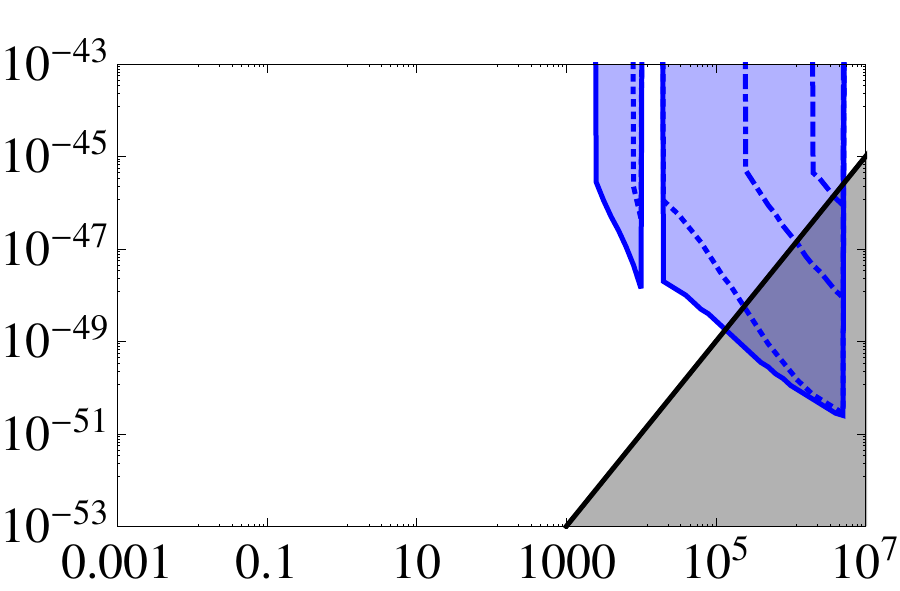} \end{minipage} &~~{\rotatebox[origin=c]{270}{$m_\phi = 50~{\rm MeV}$}}\\
&\begin{minipage}[c]{0.40\textwidth} \includegraphics[scale=.7]{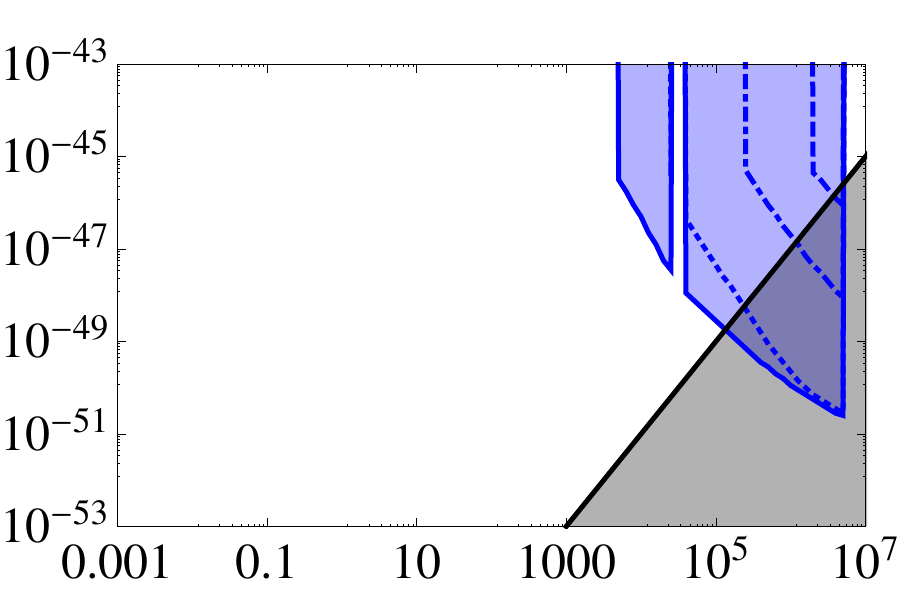} \end{minipage} &\begin{minipage}[c]{.4\textwidth}\includegraphics[scale=.7]{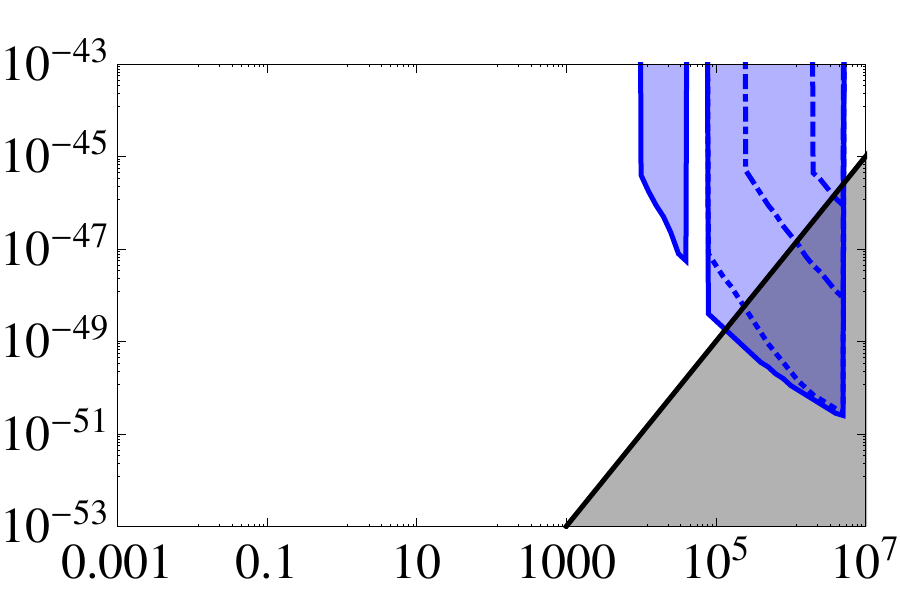} \end{minipage} &~~{\rotatebox[origin=c]{270}{$m_\phi = 100~{\rm MeV}$}}\\
&\begin{minipage}[c]{0.4\textwidth} \includegraphics[scale=.7]{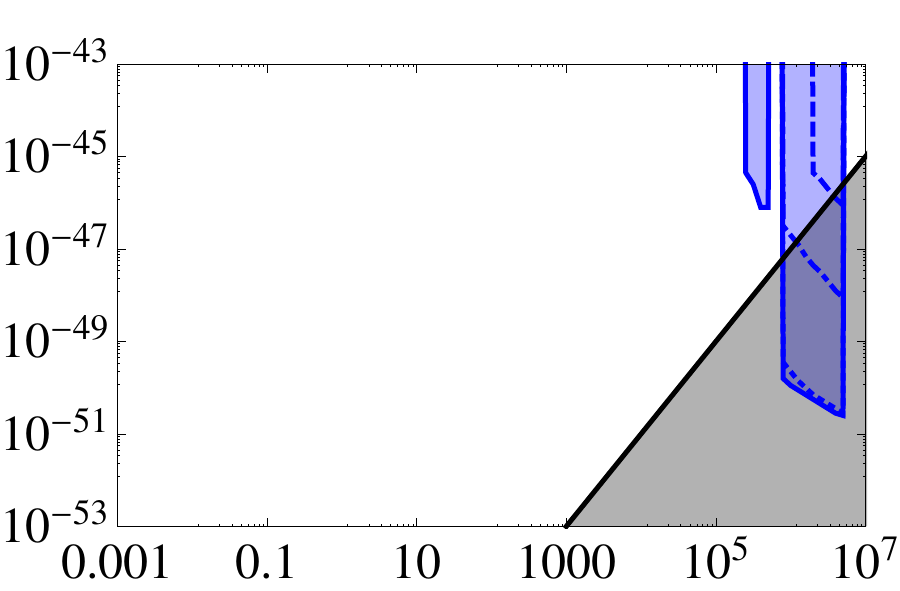} \end{minipage} &\begin{minipage}[c]{.4\textwidth}\includegraphics[scale=.7]{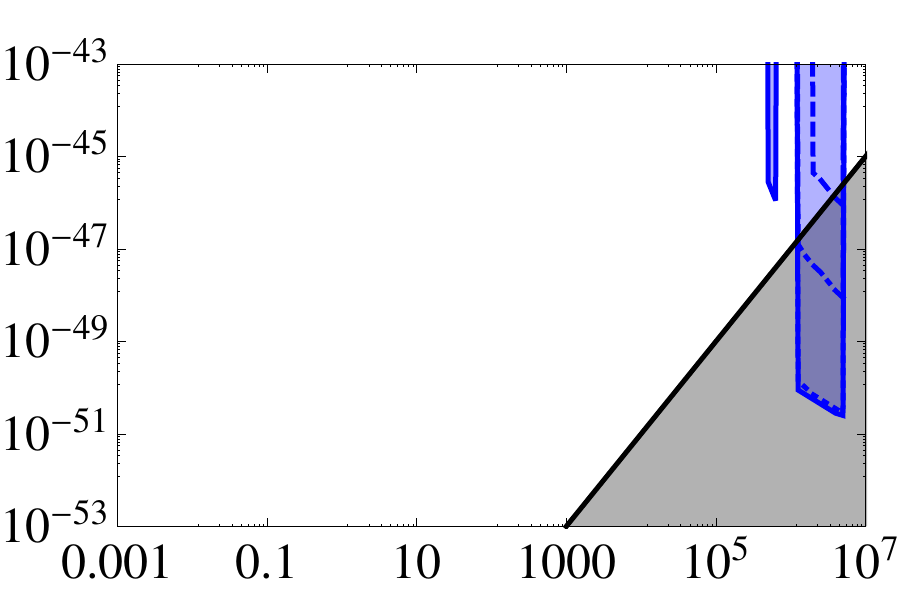} \end{minipage} &~~{\rotatebox[origin=c]{270}{$m_\phi = 500~{\rm MeV}$}}\\
 &  \multicolumn{2}{c}{Mass of Fermionic Dark Matter $m_X$ (GeV)} &
 \end{tabular}
\end{figure}

\begin{figure}
\caption{{\bf Co-annihilation:} Neutron star collapse bounds are shown in the $(m_X, \sigma_{nX})$ plane for dark matter fermions which co-annihilate with baryons and have an attractive Yukawa self-coupling. Yukawa couplings $\alpha = \{10^{-1},10^{-2} \}$ and Yukawa boson masses $m_\phi = \{50,100,500  \}~{\rm MeV}$ are displayed, and bounds on dark matter with a co-annihilation cross section of $\Expect{\sigma_a v} = \{0,10^{-55},10^{-53} \} ~{\rm cm^3/s}$ are drawn with solid, dotted, and dotted-dashed lines respectively. The shaded grey regions at the bottom right of each plot are unbounded because the dark matter has
insufficient time to thermalize, as detailed in Section \ref{sectioncollapse}.}
 \begin{tabular}{cccc}
 & \begin{minipage}[c]{0.4\textwidth} \center ~~$\alpha =10^{-1}$ \end{minipage} &\begin{minipage}[c]{0.4\textwidth} \center ~~ $\alpha =10^{-2}$ \end{minipage} & \\
{\multirow{4}{*}{\rotatebox[origin=c]{90}{Nucleon-DM Scattering Cross Section $\sigma_{nX}$ $\rm (cm^2)$}}}
&\begin{minipage}[c]{0.40\textwidth} \includegraphics[scale=.70]{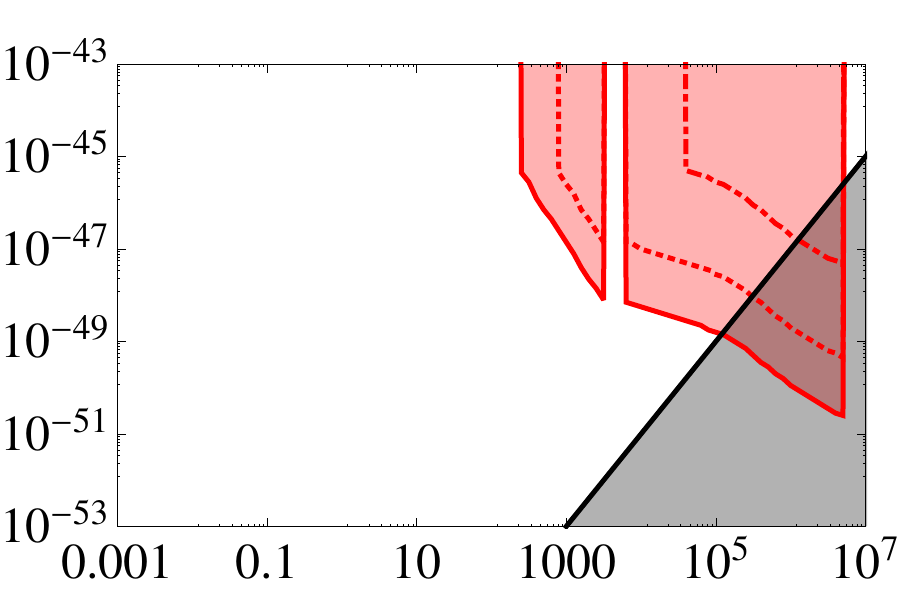} \end{minipage} &\begin{minipage}[c]{.4\textwidth}\includegraphics[scale=.70]{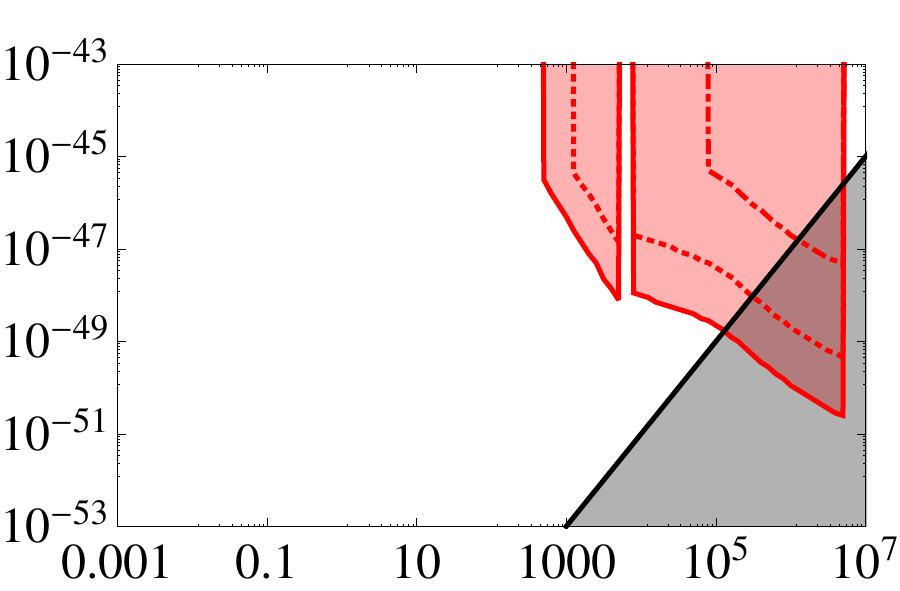} \end{minipage} &~~{\rotatebox[origin=c]{270}{$m_\phi = 50~{\rm MeV}$}}\\
&\begin{minipage}[c]{0.40\textwidth} \includegraphics[scale=.7]{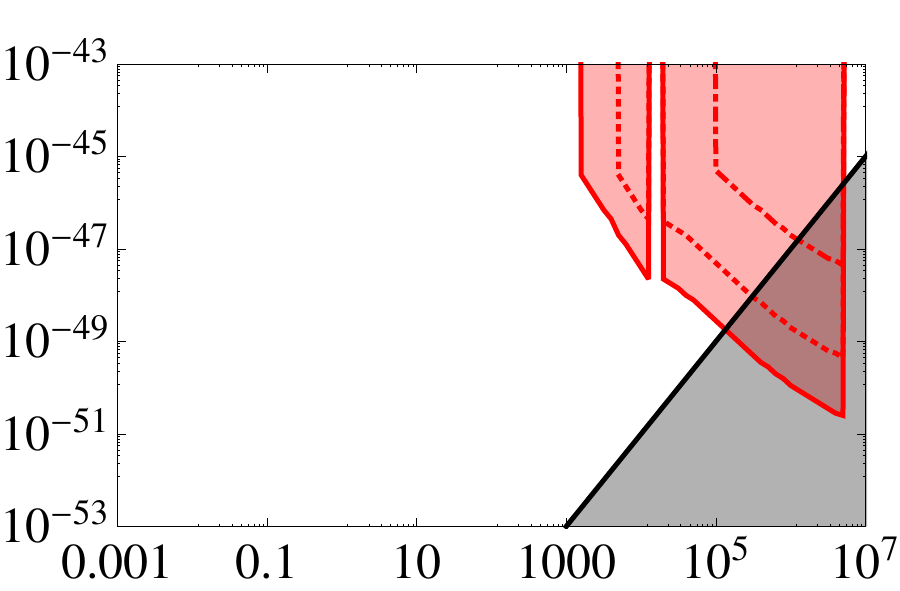} \end{minipage} &\begin{minipage}[c]{.4\textwidth}\includegraphics[scale=.7]{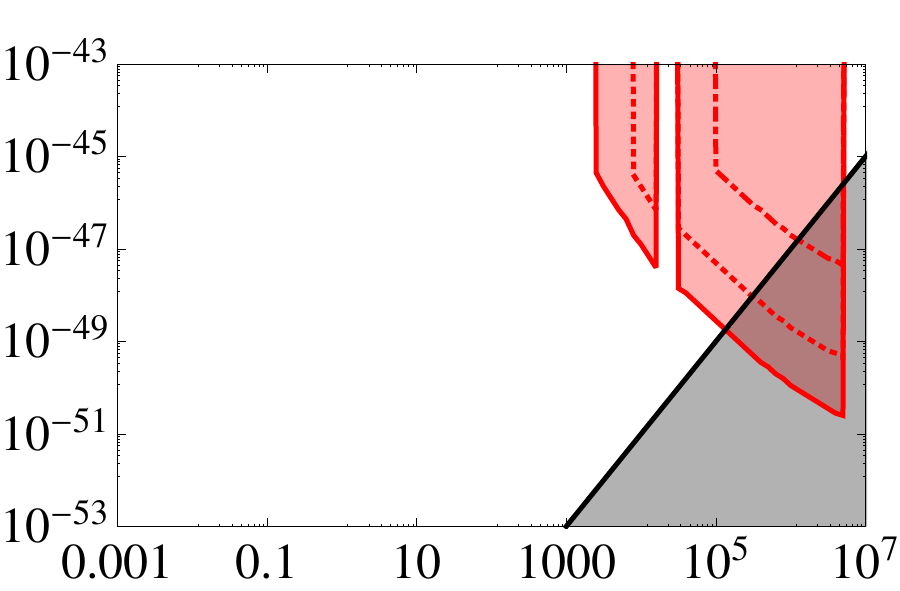} \end{minipage} &~~{\rotatebox[origin=c]{270}{$m_\phi = 100~{\rm MeV}$}}\\
&\begin{minipage}[c]{0.4\textwidth} \includegraphics[scale=.7]{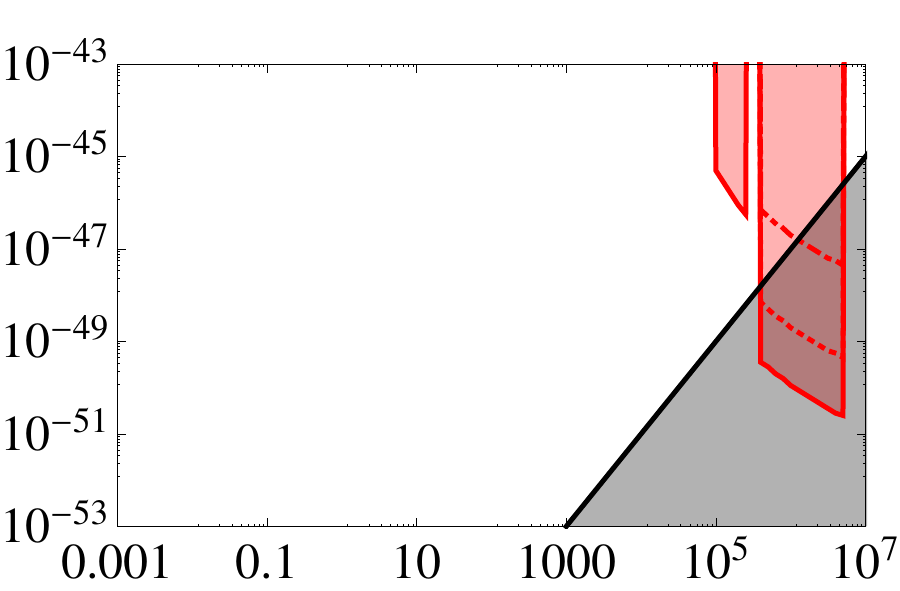} \end{minipage} &\begin{minipage}[c]{.4\textwidth}\includegraphics[scale=.7]{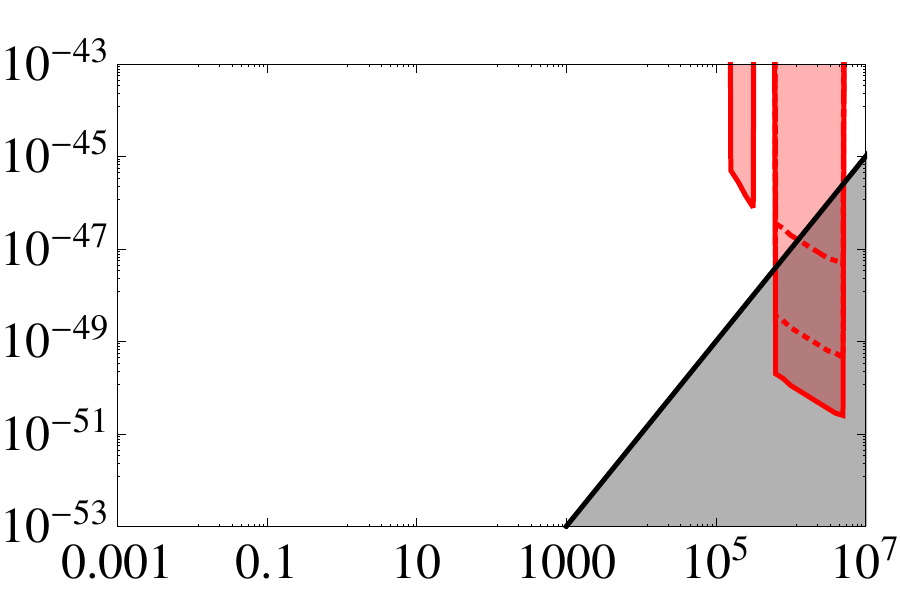} \end{minipage} &~~{\rotatebox[origin=c]{270}{$m_\phi = 500~{\rm MeV}$}}\\
 &  \multicolumn{2}{c}{Mass of Fermionic Dark Matter $m_X$ (GeV)} &
 \end{tabular}
\end{figure}

\begin{figure}
\caption{{\bf Co-annihilation:} Neutron star collapse bounds are shown in the $(m_X, \sigma_{nX})$ plane for dark matter fermions which co-annihilate with baryons and have an attractive Yukawa self-coupling. Yukawa couplings $\alpha = \{10^{-3},10^{-4} \}$ and Yukawa boson masses $m_\phi = \{1,5,10  \}~{\rm MeV}$ are displayed, and bounds on dark matter with a co-annihilation cross section of $\Expect{\sigma_a v} = \{0,10^{-55},10^{-53} \} ~{\rm cm^3/s}$ are drawn with solid, dotted, and dotted-dashed lines respectively. The shaded grey regions at the bottom right of each plot are unbounded because the dark matter has
insufficient time to thermalize, as detailed in Section \ref{sectioncollapse}.}
 \begin{tabular}{cccc}
 & \begin{minipage}[c]{0.4\textwidth} \center ~~$\alpha =10^{-3}$ \end{minipage} &\begin{minipage}[c]{0.4\textwidth} \center ~~ $\alpha =10^{-4}$ \end{minipage} & \\
{\multirow{4}{*}{\rotatebox[origin=c]{90}{Nucleon-DM Scattering Cross Section $\sigma_{nX}$ $\rm (cm^2)$}}}
&\begin{minipage}[c]{0.40\textwidth} \includegraphics[scale=.70]{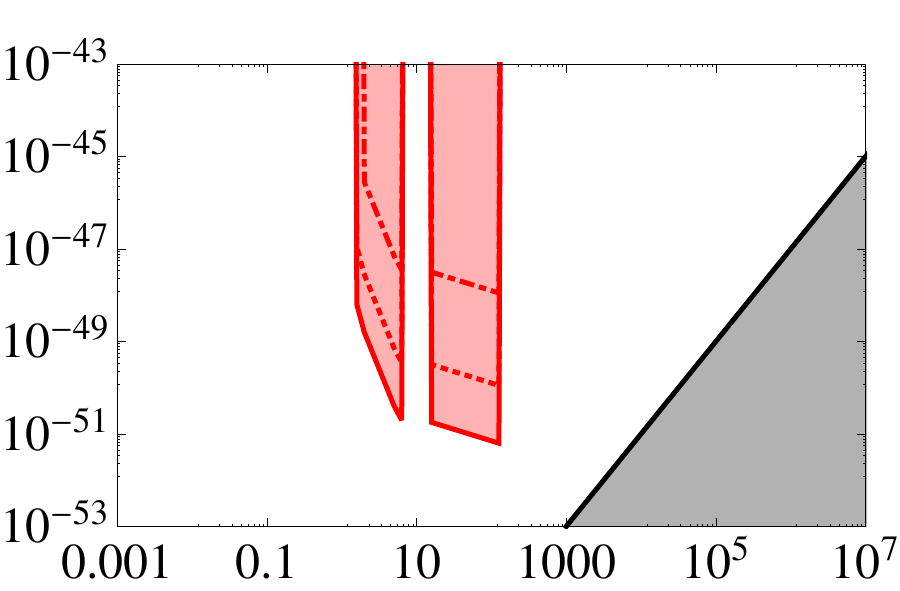} \end{minipage} &\begin{minipage}[c]{.4\textwidth}\includegraphics[scale=.70]{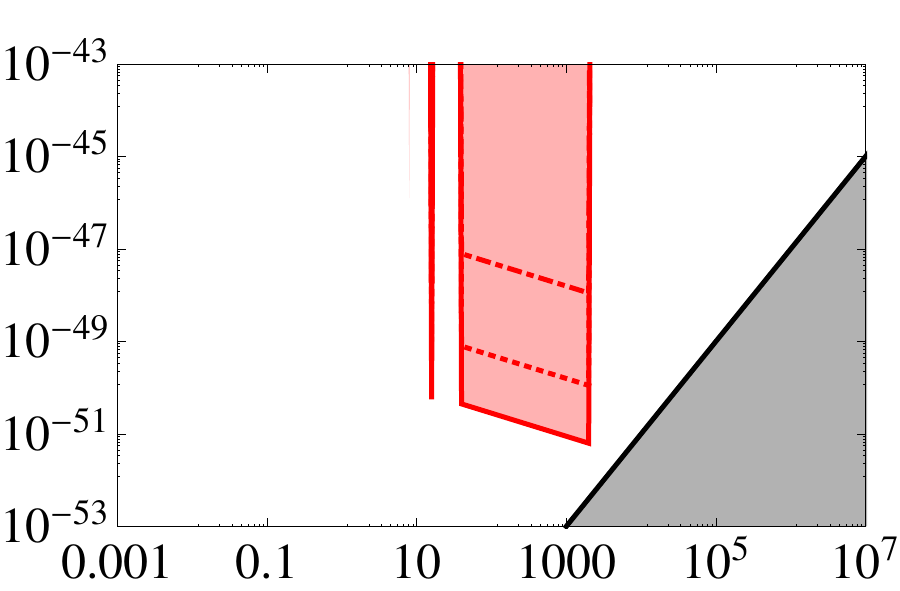} \end{minipage} &~~{\rotatebox[origin=c]{270}{$m_\phi = 1~{\rm MeV}$}}\\
&\begin{minipage}[c]{0.40\textwidth} \includegraphics[scale=.7]{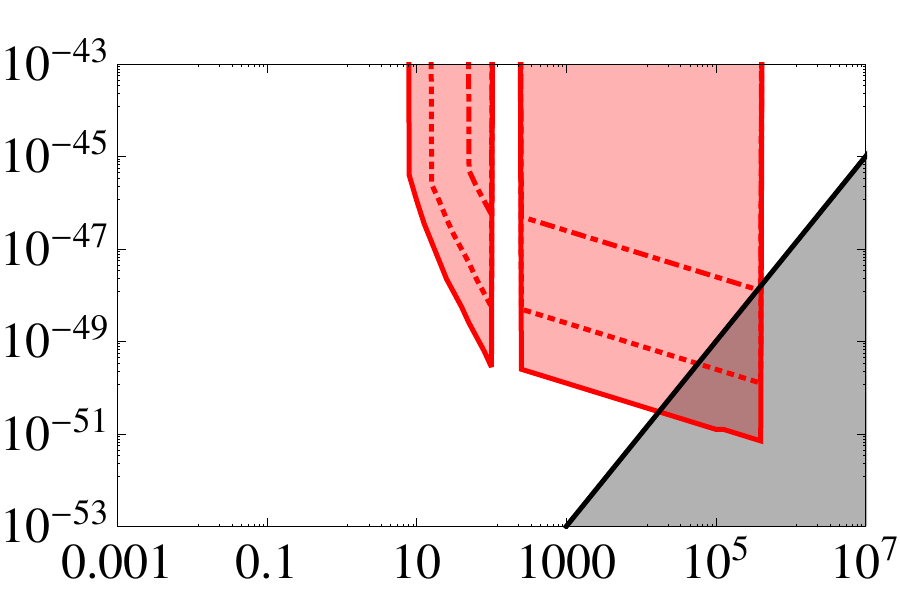} \end{minipage} &\begin{minipage}[c]{.4\textwidth}\includegraphics[scale=.7]{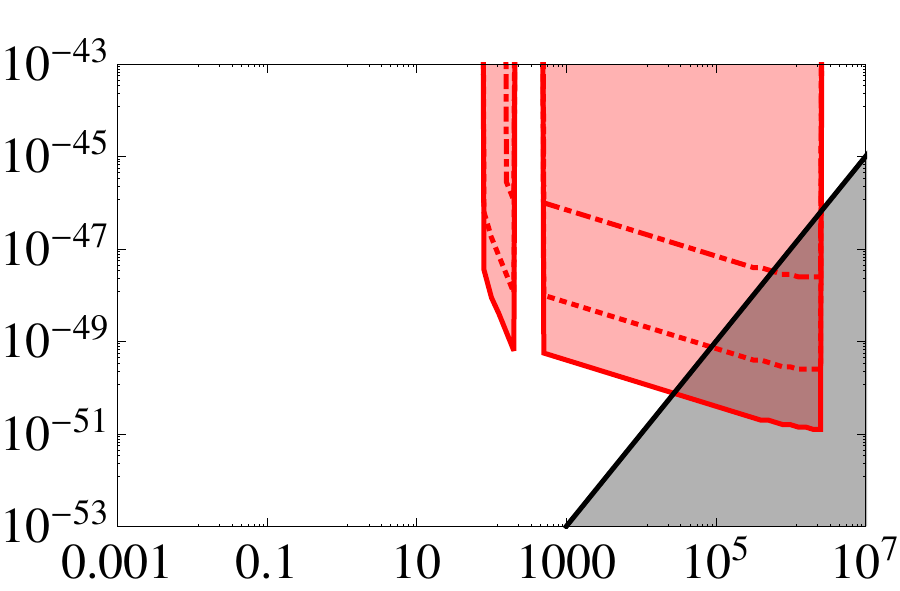} \end{minipage} &~~{\rotatebox[origin=c]{270}{$m_\phi = 5~{\rm MeV}$}}\\
&\begin{minipage}[c]{0.4\textwidth} \includegraphics[scale=.7]{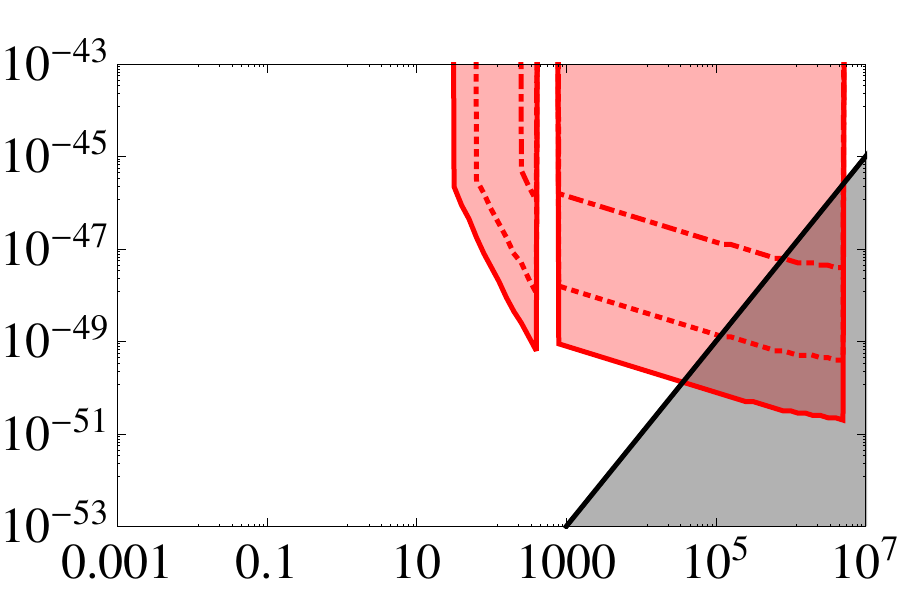} \end{minipage} &\begin{minipage}[c]{.4\textwidth}\includegraphics[scale=.7]{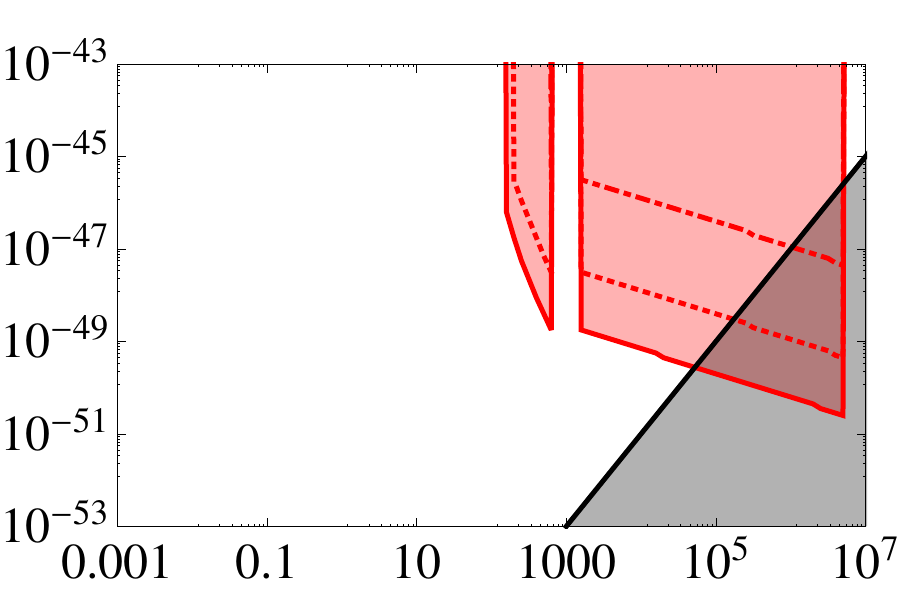} \end{minipage} &~~{\rotatebox[origin=c]{270}{$m_\phi = 10~{\rm MeV}$}}\\
 &  \multicolumn{2}{c}{Mass of Fermionic Dark Matter $m_X$ (GeV)} &
 \end{tabular}
 \label{fig7}
\end{figure}

\begin{figure}
\caption{{\bf Co-annihilation:} Neutron star collapse bounds are shown in the $(m_X, \sigma_{nX})$ plane for dark matter fermions which co-annihilate with baryons and have an attractive Yukawa self-coupling. Yukawa couplings $\alpha = \{10^{-3},10^{-4} \}$ and Yukawa boson masses $m_\phi = \{50,100,500  \}~{\rm MeV}$ are displayed, and bounds on dark matter with a co-annihilation cross section of $\Expect{\sigma_a v} = \{0,10^{-55},10^{-53} \} ~{\rm cm^3/s}$ are drawn with solid, dotted, and dotted-dashed lines respectively. The shaded grey regions at the bottom right of each plot are unbounded due to insufficient time to thermalize, as detailed in Section \ref{sectioncollapse}.}
 \begin{tabular}{cccc}
 & \begin{minipage}[c]{0.4\textwidth} \center ~~$\alpha =10^{-3}$ \end{minipage} &\begin{minipage}[c]{0.4\textwidth} \center ~~ $\alpha =10^{-4}$ \end{minipage} & \\
{\multirow{4}{*}{\rotatebox[origin=c]{90}{Nucleon-DM Scattering Cross Section $\sigma_{nX}$ $\rm (cm^2)$}}}
&\begin{minipage}[c]{0.40\textwidth} \includegraphics[scale=.70]{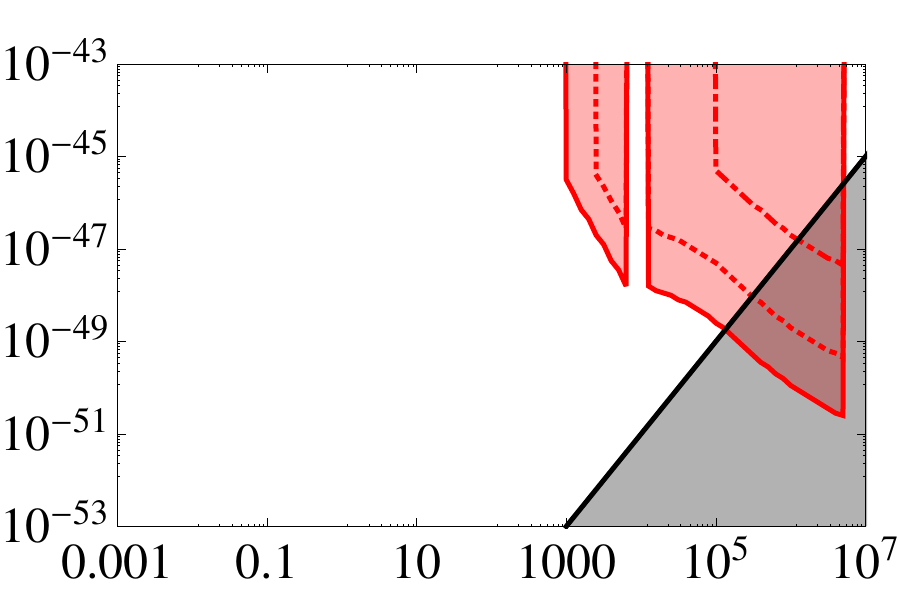} \end{minipage} &\begin{minipage}[c]{.4\textwidth}\includegraphics[scale=.70]{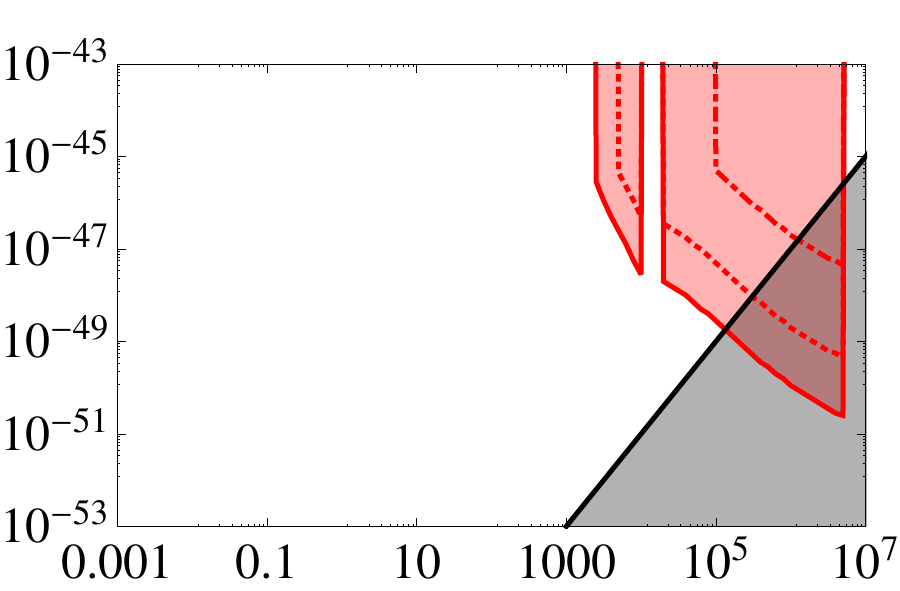} \end{minipage} &~~{\rotatebox[origin=c]{270}{$m_\phi = 50~{\rm MeV}$}}\\
&\begin{minipage}[c]{0.40\textwidth} \includegraphics[scale=.7]{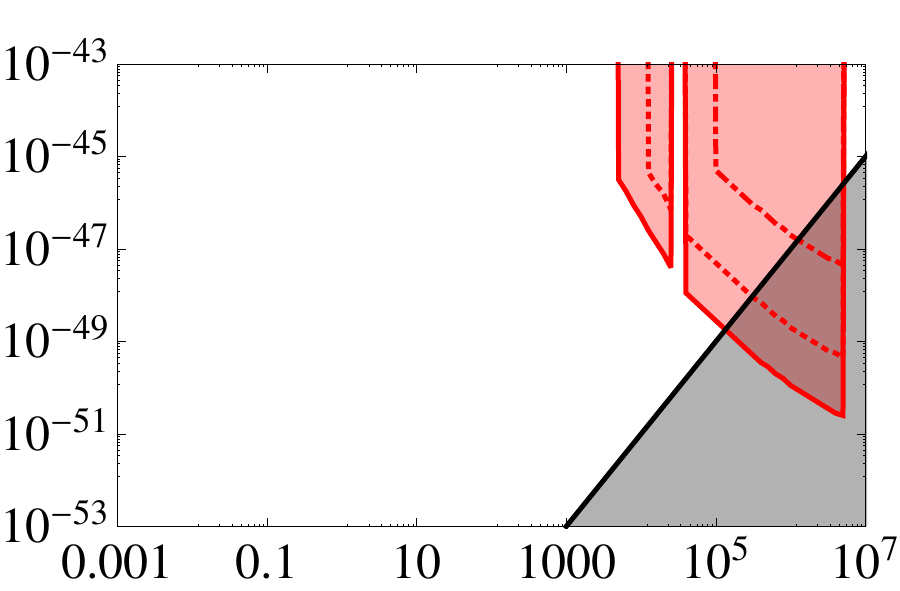} \end{minipage} &\begin{minipage}[c]{.4\textwidth}\includegraphics[scale=.7]{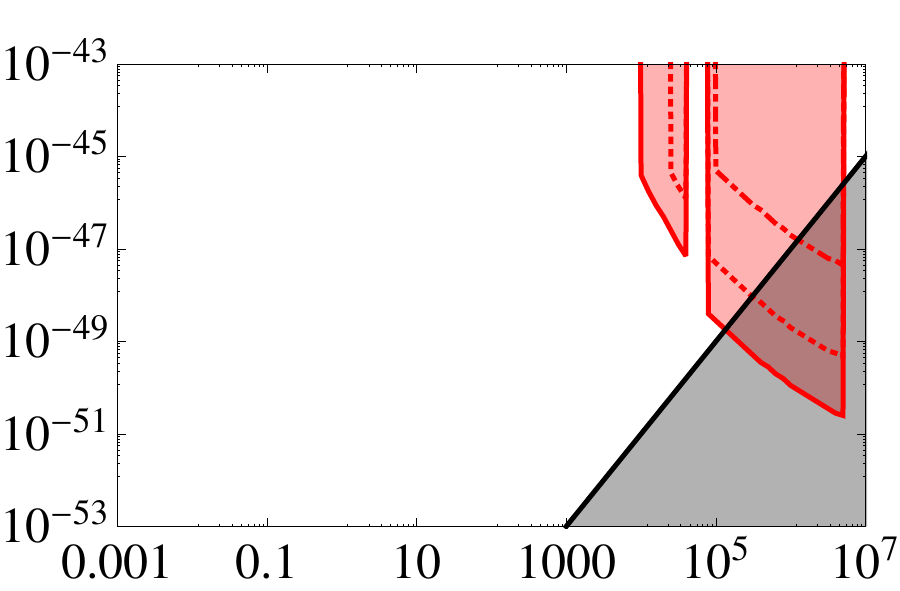} \end{minipage} &~~{\rotatebox[origin=c]{270}{$m_\phi = 100~{\rm MeV}$}}\\
&\begin{minipage}[c]{0.4\textwidth} \includegraphics[scale=.7]{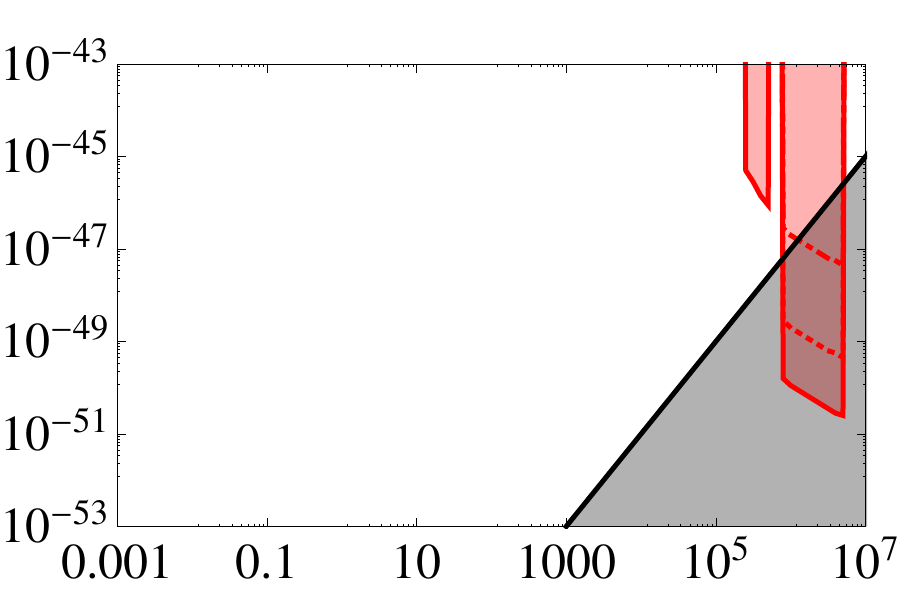} \end{minipage} &\begin{minipage}[c]{.4\textwidth}\includegraphics[scale=.7]{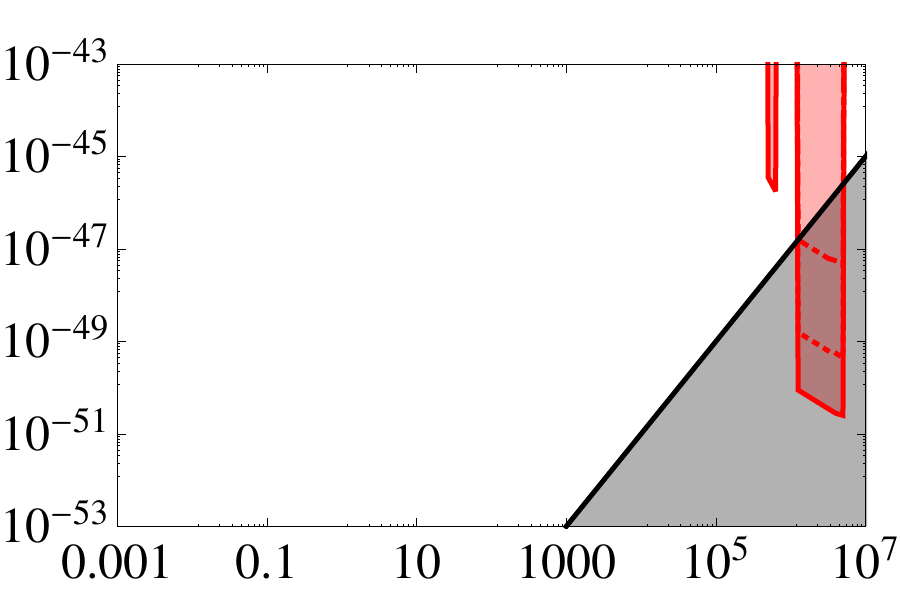} \end{minipage} &~~{\rotatebox[origin=c]{270}{$m_\phi = 500~{\rm MeV}$}}\\
 &  \multicolumn{2}{c}{Mass of Fermionic Dark Matter $m_X$ (GeV)} &
 \end{tabular}
\end{figure}
\pagebreak

\bibliographystyle{JHEP}

\bibliography{ADM}

\end{document}